# Lattice dynamics of hexagonal Zn$_{1-x}$Mg$_x$S

A. Elmahjoubi,[1,†] Mala N. Rao,[2,3] A. Ivanov,[4] A. V. Postnikov,[1] A. Polian,[5] T. Alhaddad,[1] S. L. Chaplot,[2,3] A. Piovano,[4] S. Diliberto,[6] S. Michel,[6] A. Maillard,[7] K. Strzałkowski,[8] and O. Pagès[1,*]

[1]LCP-A2MC, UR 4632, Université de Lorraine, 57000 Metz, France
[2]Solid State Physics Division, Bhabha Atomic Research Centre, Mumbai – 400085, India
[3]Homi Bhabha National Institute, Anushaktinagar, Mumbai – 400094, India
[4]Institut Laue-Langevin (ILL), 71 Avenue des Martyrs, 38042 Grenoble, France
[5]Institut de Minéralogie, de Physique des Matériaux et de Cosmochimie, Sorbonne Université - UMR CNRS 7590, 75005 Paris, France
[6]Institut Jean Lamour, Université de Lorraine - UMR CNRS 7198, 54000 Nancy, France
[7]LMOPS, Université de Lorraine - Supélec, 57070 Metz, France
[8]Institute of Physics, Faculty of Physics, Astronomy and Informatics, Nicolaus Copernicus University in Toruń, ul. Grudziądzka 5, 87-100 Toruń, Poland

## Abstract

Inelastic neutron scattering measurements on the hexagonal Zn$_{0.67}$Mg$_{0.33}$S semiconductor alloy reveal a bimodal pattern of the optical modes across the Brillouin zone, confirmed by first-principles simulations. Such modes are sensitive to the local fluctuations in the composition inherent to random Zn↔Mg alloying, distinguishing *homo* from *hetero* environments of a given bond (1-bond→2-mode), as is formalized for cubic alloys by the percolation model. The latter model thus emerges as a generic framework for systematizing the optical modes of semiconductor alloys in various crystal structures.

---

[†] abdelmajid.elmahjoubi@univ-lorraine.fr
[*] Corresponding author : olivier.pages@univ-lorraine.fr

Due to their simple structure involving a unique A↔B atom substitution on a regular and highly-symmetrical lattice, the high-entropy $A_{1-x}B_x$ pseudo-unary metallic alloys (HEA) and the disordered $A_{1-x}B_xC$ pseudo-binary semiconductor alloys with cubic structure set a benchmark to explore how physical properties are impacted by disorder [1,2] in a percolation context [3]. Notably, as a local physical property, the bond force constant governing the lattice dynamics is potentially sensitive to the alloy disorder at the ultimate atom scale of the A↔B substitution.

On this basis, the lattice dynamics of cubic HEA with a single atom per primitive cell, thus providing only acoustic phonons bearing a "cumulative" (bond-collective : multi-bond→1-mode) message about the alloy disorder, attracted considerable studies at various wavelengths spanning the Brillouin zone (BZ) by inelastic neutron scattering (INS) [4-6]. With two atoms per primitive cell, the pseudo-binary cubic SCA further exhibit optical (O) phonons due to an effective bond-stretching potentially offering a more bond-specific insight, in the spirit of a 1-bond→1-mode paradigm.

Our pioneering, and so far unique, INS study of a pseudo-binary cubic alloy, namely $Zn_{0.67}Be_{0.33}Se$ [7] with highly-mismatched physical properties exalting the lattice dynamics, e.g., bond length $L$ ($\frac{\Delta L}{L} \sim$ 9.0%), bond covalency $\alpha_c$ ($\frac{\Delta \alpha_c^3}{\alpha_c^3} \sim$ 31.0%) and substituent's covalent radius $R_c$ ($\frac{\Delta R_c}{R_c} \sim$ 13.5%) [8,9], supported by Raman scattering (RS) – see Fig. S1 with the S prefix standing for Supplementary Material reported in Ref. [10], revealed a dual O-mode for Be-Se (1-bond→2-mode). This Be-Se duo survives throughout the BZ, from quasi-infinite wavelengths near the BZ-centre, where the RS operates by force of its visible laser probe [11], through finite wavelengths achieved by INS, all the way to minimal ones of twice the lattice constant near the BZ-boundary [12].

The Be-Se dual O-mode in cubic $Zn_{0.67}Be_{0.33}Se$ was fruitfully assessed within an original percolation model (PM) [13], in which the O-modes distinguish, beyond bond species, "same" (*homo*) from "alien" (*hetero*) environments of a given bond (1-bond→2-mode [10]), and thus "see" the local fluctuations in composition inherent to random alloying. This deviates from the historical modified-random-element-isodisplacement (MREI, [14]) model in which the O-modes are "blind" to the alloy disorder, the like bonds of a given species vibrating at the same frequency at a given alloy composition (1-bond→1-mode), irrespectively of their local environment. Prior to the PM, an attempt to formalize (beyond the MREI model) a capacity of the O-modes to "see" the local bond environment was done within the cluster model, suggested for cubic [15] and hexagonal [16] alloys in the 1960's. However, the cluster model was recently ruled out as a realistic model for the paradigmatic case of cubic $GaAs_{1-x}P_x$ [17].

Successful applications of the PM to all so far explored cubic alloys, with zincblende (a non-exhaustive list is given in Ref. [18]) and diamond [19] structures suggested the model's universality. Generally, the splitting within the PM-duo increases with the mismatch in bond physical properties ($L, \alpha_c, R_c$). In this line, the MREI model represents a limit PM case for well-matched alloys, the PM-doublet degenerating into a MREI-singlet. Experimental (RS [20]) and *ab initio* (phonon dispersion [21,22]) data indeed support the MREI scheme for the well-matched cubic $Al_{1-x}Ga_xAs$ alloy ($\frac{\Delta L}{L} \sim$ 0.8%, $\frac{\Delta \alpha_c^3}{\alpha_c^3} \sim$ 4.0%, $\frac{\Delta R_c}{R_c} \sim$ 1.6%) [8] in which the lattice regularity is hardly challenged by alloying, the bond lengths remaining similar within a few percent [23].

In this work we test how the PM transfers by lowering the crystal symmetry from cubic/zincblende (zb) to hexagonal/wurtzite (w), using w-$Zn_{0.67}Mg_{0.33}S$ ($\frac{\Delta L}{L} \sim$ 4.0%, $\frac{\Delta \alpha_c^3}{\alpha_c^3} \sim$ 18.6%, $\frac{\Delta R_c}{R_c} \sim$ 15.0%) [8,9] as a case study. The selected hexagonal alloy ideally matches in composition the reference zb-$Zn_{0.67}Be_{0.33}Se$ one [7], which facilitates comparison. Our attention focuses on the non-polar O-modes – that assimilate with purely mechanical (mass + spring) oscillators – because they hardly couple and thus preserve the natural diversity of the O-modes in a complex system like an alloy [18,24].

$Zn_{1-x}Mg_xS$ is further interesting for assessing the predictive power of the PM in a non-standard situation: the lighter (Mg) substituent has the larger covalent radius in $Zn_{1-x}Mg_xS$ and hence forms the longer bond, a unique case among all alloys so far tested [18]. This invites to anticipate a PM "inversion", a situation in which the PM scheme for $Zn_{0.67}Mg_{0.33}S$ would make a fair "mirror image" in frequency and intensity of that for $Zn_{0.67}Be_{0.33}Se$ (Fig. S1). Accordingly the triple $\{O_{Zn-Se}, minor\text{-}O_{Be-Se}^{Be}, dominant\text{-}O_{Be-Se}^{Zn}\}$ signal of $Zn_{0.67}Be_{0.33}Se$ would transform into a $\{dominant\text{-}O_{Zn-S}^{Zn}, minor\text{-}O_{Zn-S}^{Mg}, O_{Mg-S}\}$ signal for $Zn_{0.67}Mg_{0.33}S$ – in this order of frequency, using the PM-notation of an optical mode ($O$) due to a given bond (subscript) in a given environment (superscript).

Preliminary backward/forward-RS studies of w-$Zn_{0.67}Mg_{0.33}S$ in the phonon/phonon-polariton regimes (Fig. S3) fail to resolve the searched $\{O_{Zn-S}^{Zn}, O_{Zn-S}^{Mg}\}$ PM-duo for Zn-S. A crude MREI-like $\{O_{Zn-S}, O_{Mg-S}\}$ signal is instead observed. Already, this deviates from a recent shell-model prediction of a mixed-mode behavior for zb-Zn$_{1-x}$Mg$_x$S [25], a rare case in which the two MREI modes merge into one. However, it is difficult to decide from the reported RS data whether the PM applies to w-Zn$_{1-x}$Mg$_x$S, or not, due to a sharp antiresonance that obscures the Zn-S signal (marked by a star, Fig. S3), as observed for the Zn-Se signal of Zn$_{1-x}$Be$_x$Se [26]. This motivates further insight beyond the BZ-centre probed by RS toward finite wavelengths using INS. Additional motivation for INS arises in that, based on the experience for Zn$_{0.67}$Be$_{0.33}$Se [7], the PM-duo seems to gain in contrast on departing from the BZ-centre. Generally, the (Zn; Mg; S) combination is well suited for INS due to their small neutron absorption cross sections $\sigma_{XS}$ = (0.9; 0.1; 0.5) barn [27] and high coherent neutron scattering lengths $b_{coh}$ = (5.2; 5.7; 2.8) fm [28].

In this letter, a dispersion of the {$O^{Zn}_{Zn-S}$, $O^{Mg}_{Zn-S}$} duo within non-polar O-modes of w-Zn$_{0.67}$Mg$_{0.33}$S is searched for by INS across the BZ along the $\overrightarrow{c^*}$ and $\overrightarrow{a^*}$ high-symmetry reciprocal lattice directions. An additional w-Zn$_{0.94}$Mg$_{0.06}$S alloy with minimal Mg content preserving the hexagonal structure is used as a "negative" PM-control, meaning that the PM-duo should not show up in this case (because alloying tends to zero). The discussion of the INS data is supported by *ab initio* phonon calculations (using the SIESTA code [29]) done on large (up to 360-atom) disordered w-Zn$_{1-x}$Mg$_x$S supercells at high ($x$=0.33, Fig. S4) and low ($x$~0,1) alloying. The *ab initio* insight is used as a "positive" PM-control – meaning that any presumed PM-duo seen by INS should replicate *ab initio* if intrinsic to random Zn↔Mg alloying ($x$=0.33), and also to track down a possible origin of the PM-splitting in w-Zn$_{1-x}$Mg$_x$S at the microscopic scale ($x$~0,1).

Our ambition is twofold: first, to achieve a unified understanding of the lattice dynamics of alloys under the shift of paradigm from the MREI to the PM, beyond cubic and onto hexagonal crystal structures, and, second, to set the PM as a robust tool able to manage bond specificities (related to $R_c$ and $L$ in this case). Moreover, as we are not aware of any INS study of a hexagonal alloy, the current INS experiment planned for w-Zn$_{1-x}$Mg$_x$S is interesting *per se*.

The INS experiment on Zn$_{1-x}$Mg$_x$S is carried out on two large (Fig. S2) homogenous (checked by Raman mapping across the ingots) free-standing single crystals with wurtzite structure verified by powder X-ray diffraction, grown using the Bridgman method [30] at high (Zn$_{0.67}$Mg$_{0.33}$S, half-cylinder 15 × 8 mm$^2$ in length×diameter) and low (Zn$_{0.94}$Mg$_{0.06}$S, cylinder 6×8 mm$^2$) alloying. High quality of crystal structure is attested by sharp peaks in the powder X-ray diffractograms (Fig. S2). The single crystals are oriented by conoscopy ($\overrightarrow{c^*}$ axis), X-ray diffraction and neutron diffraction *in situ* ($\overrightarrow{c^*}$ and $\overrightarrow{a^*}$ axis). The composition $x$ is determined with accuracy 0.1% by the inductively coupled plasma method.

INS experiments are conducted at the thermal neutron three-axis spectrometer IN8 [31] at the high flux reactor of the Institut Laue-Langevin. The 2D-focusing copper monochromator with reflection planes Cu-(200) and pyrolytic graphite (PG) analyser with reflection planes PG002 provide optimal spectral resolution of the O-modes using scattered neutron wave vector 2.662 Å$^{-1}$. Higher order contaminations in the monochromatic beams were practically entirely removed with a 10 cm thick oriented pyrolytic graphite filter. The INS spectra are measured at sample temperature of 12 K in the $\overrightarrow{c^*}/\overrightarrow{a^*}$ scattering plane at various phonon wave vectors across the BZ along $\overrightarrow{c^*}$ and $\overrightarrow{a^*}$. Both the acoustic and optical modes are probed for Zn$_{0.67}$Mg$_{0.33}$S, whereas only the non-polar O-modes, of main interest, for Zn$_{0.94}$Mg$_{0.06}$S. Experiment was guided by dynamical structure factor computations using the shell model [25], that helps to target special BZ in the periodic zone scheme (shifted from the first BZ by a reciprocal lattice vector $\vec{G}$) as the best compromises between maximal neutron scattering cross section and optimal symmetry (transverse or longitudinal) access, taking into account a selection rule that only the modes polarized along the transferred wavevector $\vec{Q} = \vec{G} + \vec{q}$ are detected by INS [12].

*Ab initio* phonon dispersions of w-Zn$_{0.67}$Mg$_{0.33}$S are calculated with the SIESTA code based on the density functional theory (DFT), using specifically the local density approximation (LDA), by the *finite displacement* technique. Two large (360-atoms) special quasirandom structures [32] are used, optimized to a random Zn↔Mg substitution with the help of the Alloy Theoretic Automated Toolkit (ATAT [33]). The supercells are elongated in a sequence of ten base-slices cut normal to [001] and [100] directions (Fig. S4). Each slice is one primitive cell thick and has lateral size of 3×3 primitive cells. Once the *finite displacement* calculation is performed for a supercell, different projections of the set of 3N (N: number of atoms) vibration eigenvectors are applied in order to reveal different trends – as explained in Sec. N.3 of Ref. [29]. In this work, combined projections onto $\vec{q}$ wavevectors and onto irreducible presentations of the wurtzite space group enabled extraction of the "spectral functions", resolved in frequency and wavevector, to be directly confronted with the INS w-Zn$_{0.67}$Mg$_{0.33}$S phonon dispersions. For practical reasons, the "sharpest" projections of this kind come out for $\vec{q}$ wavevectors being divisions of the reciprocal lattice vector by the supercell multiplication factor. Apart from this, simulations at the onset of the PM-splitting of the non-polar O-modes have been done on 72-atom supercells containing one isolated Mg or Zn impurity.

*Quadruple test on the non-polar O-modes.* – Four atoms per primitive cell in the wurtzite structure generate 3×4=12 modes which split into the following irreducible representations at the BZ-centre : $2A_1 \oplus 2B_1 \oplus 2E_1 \oplus 2E_2$ [34]. Out of these, one set of $A_1$ and one set of $E_1$ are acoustic modes, the remaining $A_1 \oplus 2B_1 \oplus E_1 \oplus 2E_2$ are O-modes, polarized along the $\vec{c}$-axis ($A_1$, $B_1$) or perpendicular to it ($E_1$, $E_2$), as sketched out in Fig. S3. In the following we refer abusively to the individual phonon modes across the BZ using their irreducible representation at the BZ-centre, discriminating low ($L$) and high ($H$) vibrations within both $B_1$ and $E_2$.

$B_1$ and $E_2$ are non-polar, hence potentially interesting candidates for the test of the PM on w-Zn$_{1-x}$Mg$_x$S (abbreviated PM-test). The low-frequency $B_1^L$ and $E_2^L$ modes are bond-insensitive in that the two bonds vibrate at the same frequency (bond-collective), hence *a fortiori* environment-insensitive, and thus irrelevant for the PM-test. The high-frequency bond-specific $B_1^H$ and $E_2^H$ modes, on the contrary, are retained. $A_1$ and $E_1$ create a finite dipolar moment *per* primitive cell, hence are polar in character. As such, they exhibit a phonon-polariton dispersion within the first $10^{-4}$ of the BZ size, governed by the polar LO asymptote near the BZ-centre and by the non-polar TO asymptote away from it [11]. Only the latter non-polar $A_1(TO)$ and $E_1(TO)$ asymptotic modes, probed in backward-RS and by INS near the BZ-centre, are relevant for the PM-test. Altogether, this leaves four non-polar O-modes for the PM-test, *i.e.*, $A_1(TO)$, $E_1(TO)$, $B_1^H$ and $E_2^H$. The intended PM-test should comprise all the four and will be decisive, in either positive or negative sense, only if the four would respond in the same way.

*Dual ab initio insight at x~(0,1).* – Long Mg-S impurity bonds created by a unique Mg↔Zn substitution in a 72-atom w-ZnS supercell ($x$~0, Fig. S5) produce a compressive strain in the ZnS-like host matrix formed with short bonds. The few Zn-S bonds in the *hetero* environment of Mg are thus shorter than the numerous matrix-like Zn-S bonds vibrating in *homo* environment away from Mg, *i.e.*, $L_{Zn-S}^{Mg} < L_{Zn-S}^{Zn}$, in a notation previously used for the O-modes, employing the length $L$ in place of TO. This results in a bimodal (minor-$L_{Zn-S}^{Mg}$, dominant-$L_{Zn-S}^{Zn}$) distribution of Zn-S bond lengths, characterized by large mismatches "along" $\vec{c}$ ($\Delta L_{Zn-S}^{\|\vec{c}}$) and "perpendicular" to $\vec{c}$ ($\Delta L_{Zn-S}^{\perp\vec{c}}$). This generates in turn a dual {dominant-$O_{Zn-S}^{Zn}$, minor-$O_{Zn-S}^{Mg}$} INS signal for Zn-S with a large frequency gap ($\delta\omega_{Zn-S}^{\|\vec{c},\perp\vec{c}}$~4.10 meV) in all {$A_1(TO)$, $E_1(TO)$, $B_1^H$, $E_2^H$} symmetries. The same *ab initio* protocol at $x$~1 (Fig. S6) gives a compact {minor-$O_{Mg-S}^{Zn}$, dominant-$O_{Mg-S}^{Mg}$} Mg-S duo, split by $\delta\omega_{Mg-S}^{\perp\vec{c}}$~2.83 meV for $E_1(TO)$ and $E_2^H$ polarized "transverse" to $\vec{c}$ (superscript), and by merely $\delta\omega_{Mg-S}^{\|\vec{c}}$~ 1.51 meV for $A_1(TO)$ and $B_1^H$ polarized "along" $\vec{c}$.

$\delta\omega_{Zn-S}$ exceeds the full width at half maximum of most INS lines (~1.50 meV) so that the Zn-S PM-duo should be resolved by INS in any {$A_1(TO)$, $E_1(TO)$, $B_1^H$, $E_2^H$} symmetry. For the compact Mg-S PM-duo, this is not so obvious for $E_1(TO)$ and $E_2^H$ and quasi hopeless for $A_1(TO)$ and $B_1^H$. Generally, $\delta\omega_{Mg-S} < \delta\omega_{Zn-S}$ reflects a basic rule within the PM that $\delta\omega$ is large/small for the small/large-$R_c$ substituent. Indeed, the small-Zn/large-Mg substituent has more/less room in its S-cage to accommodate locally the ($\alpha_c$, $L$)-bond mismatch. So, its phonon frequencies are more/less diversified.

*Triple $2\times$ experimental ($x$=0.06; 0.33) vs. $1\times$ ab initio insight ($x$=0.33).* – An overview of the entire $2A_1 \oplus 2B_1 \oplus 2E_1 \oplus 2E_2$ phonon dispersions of w-Zn$_{0.67}$Mg$_{0.33}$S measured by INS across the BZ along $\vec{c^*}$ and $\vec{a^*}$ is juxtaposed with corresponding *ab initio* dispersions in Fig. 1, for comparison. If we omit a slight shift between experimental and *ab initio* frequencies (less than 1.85 meV) due to a well-known LDA bias overestimating the force constants [29], the experimental and *ab initio* dispersions are remarkably consistent. In particular, the bond-specific $A_1(LO) - B_1^H$ and $E_1(TO) - E_2^H$ and the bond-collective $A_1(LA) - B_1^L$ and $E_1(TA) - E_2^L$ degeneracies at the BZ-boundary along $\vec{c^*}$, visible with w-ZnS [35] and with w-MgS [36,37], are attested for w-Zn$_{0.67}$Mg$_{0.33}$S, both in experiment and *ab initio*.

The PM-test focuses on the non-polar {$A_1(TO)$, $E_1(TO)$, $B_1^H$, $E_2^H$} O-modes, as already discussed. The first three are probed in their dispersion along $\vec{c^*}$ and the fourth one along $\vec{a^*}$. The $E_1(TO)$ (Fig. 2), $E_2^H$ (Fig. S7), $A_1(TO)$ (Fig. S8) and $B_1^H$ (Fig. S9) dispersions measured by INS at high alloying on w-Zn$_{0.67}$Mg$_{0.33}$S (Fig. 2a) are discussed in comparison with their counterparts calculated *ab initio* on large special quasirandom structures (Fig. 2b), used as "positive" PM-controls (as defined above), and in the light of additional dispersions measured by INS at low alloying on w-Zn$_{0.94}$Mg$_{0.06}$S (Fig. 2c), used as "negative" PM-controls.

As soon as departing from the BZ-centre (especially sensitive for INS/*ab initio* [10]), a triple {dominant-$O_{Zn-S}^{Zn}$, minor-$O_{Zn-S}^{Mg}$, $O_{Mg-S}$} INS signal for w-Zn$_{0.67}$Mg$_{0.33}$S, including the searched for Zn-S PM-duo, scaling 2:1 in intensity, shows up in all tested symmetries across the BZ (Fig. 2a). This triple INS signal is the exact mirror-image of the zb-Zn$_{0.67}$Be$_{0.33}$Se RS signal, as anticipated (Fig. S1). In any symmetry the Zn-S duo seen by INS is replicated *ab initio* (Fig. 2b); hence the effect is intrinsic to random Zn↔Mg alloying. Last, within the Zn-S PM-duo only $O_{Zn-S}^{Zn}$ survives in the INS signal of w-Zn$_{0.94}$Mg$_{0.06}$S; $O_{Zn-S}^{Mg}$ has disappeared along with $O_{Mg-S}$ (Fig. 2c). This is because the Mg-environment shrinks to zero (impacting $O_{Zn-S}^{Mg}$) when the Mg-S bond fraction tends to zero (impacting $O_{Mg-S}$). As an added bonus, the compact PM-duo predicted *ab initio* for Mg-S in w-Zn$_{1-x}$Mg$_x$S at $x$~1 is observed at $x$~0.33 by INS near the BZ-boundary and also *ab initio* across the BZ in all relevant symmetries (*e.g.*, in Fig. 2) [10]. The Zn-S and Mg-S PM-duos are emphasized by shaded areas in Fig. 1.

Besides, the INS/*ab initio* data reveal that the acoustic modes do not become fatally overdamped near the BZ-boundary [10], at variance with observations on cubic HEA [4-6] and cubic Zn$_{0.67}$Be$_{0.33}$Se [7]. Along $\vec{c^*}$, when the acoustic modes achieve degeneracy with O-modes of same polarization (corresponding to atom displacements predominantly parallel or perpendicular to $\vec{c^*}$, Fig. S3) the inclination of the O-modes to gain in resolution near the BZ-boundary ([10] – Fig. S10), as observed for the PM-duos, takes over (Fig. 1, shaded circles). Along $\vec{a^*}$ the overdamping occurs near the BZ-boundary ([10] – Fig. S11) because the acoustic modes cannot couple with the O-modes nearby, of different polarization (Fig. 1, hollow circles). Similarly, the acoustic modes do not become fatally overdamped on crossing the optical modes halfway their dispersion ([10] – Fig. S10), as recently observed in a disordered cubic ceramics [38]. The latter modes either couple ([10] – Figs. S11 and S12), manifested by a repulsion of the coupled features [10], noted (±) in order of increasing frequency, or not ([10] – Fig. S10), depending on whether they exhibit similar (along $\vec{a^*}$) or different (along $\vec{c^*}$) polarizations.

Summarizing, our multi-symmetry (Figs. 2 and S7-to-S9) and multi-sample ($x$=0.06, 0.33) three-pronged (two experimental plus one *ab initio* insights *per* O-mode) approach ascertains that the Zn-S duo observed for all non-polar O-modes by INS on hexagonal Zn$_{0.67}$Mg$_{0.33}$S, that exhibits an odd bond mismatch, is due to alloying (present for Zn$_{0.67}$Mg$_{0.33}$S and absent for Zn$_{0.94}$Mg$_{0.06}$S) and is intrinsic to random Zn↔Mg alloying (present for Zn$_{0.67}$Mg$_{0.33}$S in experiment and also *ab initio* using special quasirandom structures). The PM-test on hexagonal Zn$_{0.67}$Mg$_{0.33}$S is thus positive. This establishes the transferability of the PM across crystal structures.

Generally, by describing an alloy in terms of a *homo*/*hetero*-composite, the PM paves the way for exploring its response to various stimuli (hydrostatic pressure – under current examination [18,39], uniaxial/biaxial stress, temperature, electric/magnetic field) at the unusual mesocopic scale, via phonons.

**Data availability statement**

The reported experimental (Raman, X-ray diffraction, inelastic neutron scattering) and *ab initio* (SIESTA code) data in this work are available upon request to the corresponding author.

**Acknowledgements**

We acknowledge assistance from the IN8 beamline staff of Institut Laue Langevin (Proposal 7-01-598 "Percolation scenario for hexagonal semiconductor alloys : A test case by applying inelastic neutron scattering to wurtzite $Zn_{0.67}Mg_{0.33}S$"), from the IJL core facility (Université de Lorraine – http://ijl.univ-lorraine.fr/recherche/centres-de-competences/rayons-x-et-spectroscopie-moessbauer-x-gamma) for the X-ray diffraction measurements, and from Pascal Franchetti in the Raman measurements. The mesocenter of calculation EXPLOR at the Université de Lorraine (Project No. 2019CPMXX0918) enabled the execution of *ab initio* calculations. This work was supported by the French PIA project «Lorraine Université d'Excellence», part of the France 2030 Program, reference ANR-15-IDEX-04-LUE, within the ViSA – IRP *«**Vi**brations of **S**emiconductor **A**lloys – **I**nternational **R**esearch **P**artnership»* wall-less associated international laboratory (2024 – 2028), co-funded by the *«Excellence Initiative – Research University program at Nicolaus Copernicus University in Toruń»*.

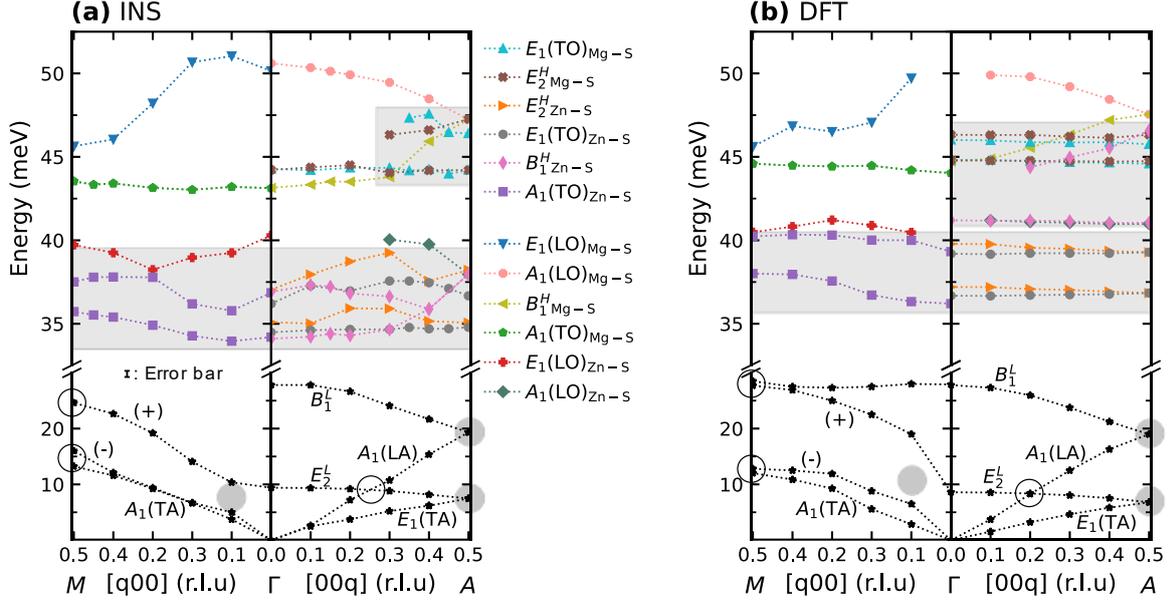

FIG. 1. Wurtzite (w) Zn$_{0.67}$Mg$_{0.33}$S phonon dispersion. (a) Experimental (INS) and (b) *ab initio* (DFT) w-Zn$_{0.67}$Mg$_{0.33}$S phonon dispersions along $\vec{a^*}$ and $\vec{c^*}$ sampled using reciprocal lattice units (r.l.u.). Dotted curves are guide for the eye. The experimental (matching the INS resolution) and *ab initio* (due to uncertainty on peak positions) error bars are within the symbol size. The shaded rectangles emphasize Zn-S and/or Mg-S PM-duos for the top-listed O-modes. The shaded/hollow circles emphasize the presence/absence of coupling on the crossing of acoustic and optical modes with same/different polarizations. The coupled modes are symbolized ± in order of frequency.

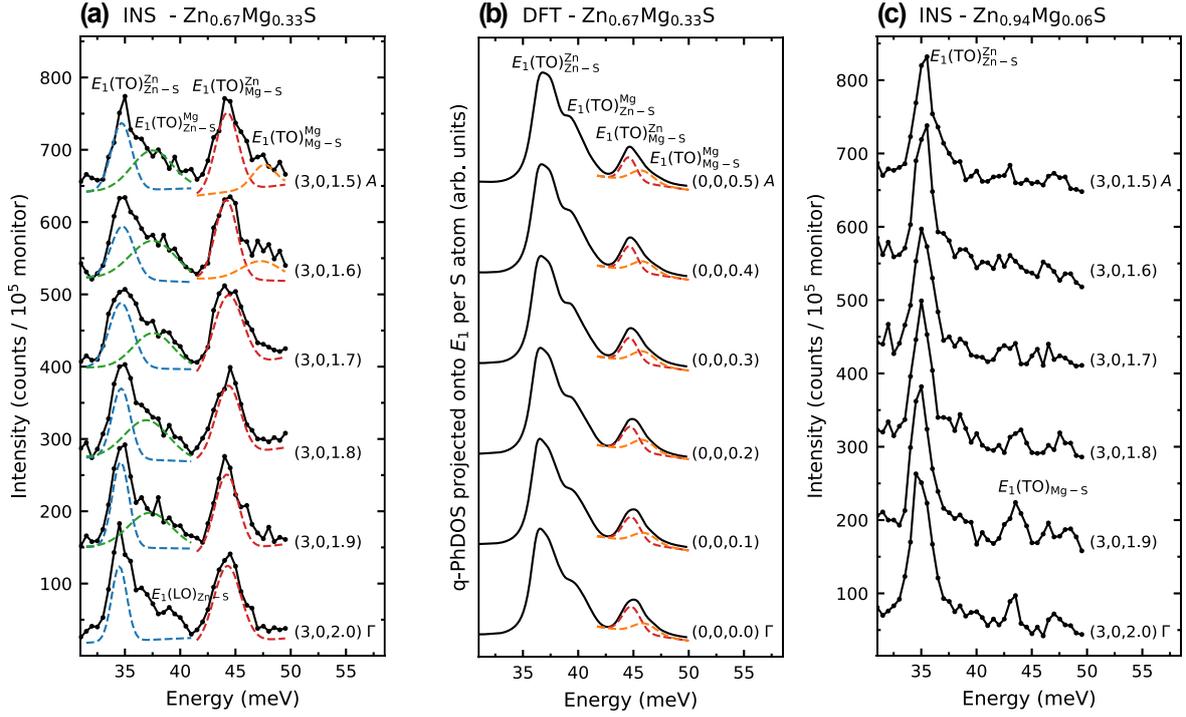

FIG. 2. $E_1(TO)$ wurtzite (w) $Zn_{1-x}Mg_xS$ phonon dispersion. (a) $E_1(TO)$ w-$Zn_{0.67}Mg_{0.33}S$ phonon dispersion measured by INS along $\vec{c^*}$. The dual Zn-S and Mg-S signals ($TO$-subscript) are deconvolved (colored dashed lines), distinguishing *homo* from *hetero* environments ($TO$-superscript). $E_1(LO)$ is activated specifically at the Brillouin zone centre, as indicated. (b) Corresponding dispersion calculated a*b initio* (DFT) per S atom. (c) The same dispersion measured by INS with w-$Zn_{0.94}Mg_{0.06}S$.

# SUPPLEMENTARY MATERIAL

# Lattice dynamics of hexagonal $Zn_{1-x}Mg_xS$

A. Elmahjoubi,[1,†] M. Rao,[2] A. Ivanov,[3] A.V. Postnikov,[1] A. Polian,[4] T. Alhaddad,[1] S.L. Chaplot,[2,3] A. Piovano,[3] S. Diliberto,[5] S. Michel,[5] A. Maillard,[6] K. Strzałkowski,[7] and O. Pagès[1,*]

[1]LCP-A2MC, UR 4632, Université de Lorraine, 57000 Metz, France
[2]Solid State Physics Division, Bhabha Atomic Research Centre, 400085 Mumbai, India
[3]Institut Laue-Langevin, 38042 Grenoble, France
[4]Institut de Minéralogie, de Physique des Matériaux et de Cosmochimie, Sorbonne Université - UMR CNRS 7590, 75005 Paris, France
[5]Institut Jean Lamour, Université de Lorraine - UMR CNRS 7198, 54000 Nancy, France
[6]LMOPS, Université de Lorraine - Supélec, 57070 Metz, France
[7] Institute of Physics, Faculty of Physics, Astronomy and Informatics, Nicolaus Copernicus University in Toruń, ul. Grudziądzka 5, 87-100 Toruń, Poland

Supplementary material (prefixed S) is concerned with the following. Sec. S1 introduces the Raman scattering (RS) signal due to the non-polar transverse optical (TO) modes of $Zn_{0.67}Be_{0.33}Se$ with zincblende (zb) structure, constituting a natural reference to discuss the RS and inelastic neutron scattering (INS) signals of the composition-matched wurtzite (w) $Zn_{0.67}Mg_{0.33}S$ semiconductor alloy of current interest. This is an occasion to briefly recall the main features of the percolation model (PM). Sec. 2 is concerned with w-$Zn_{1-x}Mg_xS$. Its INS study reported in the main text is grounded in a preliminary backward/forward RS study at high-alloying ($x$=0.33) in Sec. S2.1, with an emphasis on the optical (O) modes of relevance for the PM-test (specified via their vibration patterns). Sec. S2.2 reports on various w-$Zn_{1-x}Mg_xS$ ($x$=0.06, 0.33) phonon dispersions measured by INS across the Brillouin zone (BZ), discussed in the light of corresponding *ab initio* calculations. Preliminary *ab initio* insights at the BZ-centre using large ZnS- and MgS-like wurtzite supercells containing an isolated Mg or Zn atom, respectively, provided in Sec. S2.2.a., help to trace back to the microscopic scale the origin of a PM-type fine structuring of the Zn-S and Mg-S O-mode of w-$Zn_{1-x}Mg_xS$ – if any. The dispersions of the high-frequency bond-specific optical (O) modes, of main interest in this work, are addressed in Secs. S2.2.b. The INS and *ab initio* data related to the O-modes are compared in the format of a triptych, as adopted in the main text for $E_1(TO)$ – see Fig. 2, to facilitate comparison between phonon symmetries. Sec. S2.2.c. is concerned with the low-frequency bond-collective acoustic and optical modes, of side interest in this work. Sec. 2.2.c.1 puts an emphasis on the phonon behavior near the BZ-boundary. Sec. S2.2.c.2 reports on the apparent coupling between the optical $E_2^L$ and acoustic $A_1(LA)$ modes halfway their dispersion along $\vec{a^*}$, currently described in terms of a simple model of two mechanically-coupled harmonic oscillators in the linear chain approximation.

1. **Zincblende (zb) $Zn_{0.67}Be_{0.33}Se$**

In preamble to the forthcoming discussion of the vibrational properties of the w-$Zn_{0.67}Mg_{0.33}S$ alloy supported by RS, INS and *ab initio* phonon calculations, we find it useful to briefly introduce the RS signal of the composition-matched zb-$Zn_{0.67}Be_{0.33}Se$ alloy, for reference purpose. This is also useful to recall the "PM-machinery" in main lines.

In the 1990's, the emergence of zincblende $Zn_{1-x}Be_x$-chalcogenides involving the light Be element with small covalent radius created a new context for studying the lattice dynamics of zincblende alloys. Such element forms short and covalent/stiff bonds contrasting with the comparatively long and ionic/soft Zn-based bonds. The highly-mismatched bond physical properties clarify the vibrational properties. Our pioneering RS studies on

---

[†] Contact : abdelmajid.elmahjoubi@univ-lorraine.fr
[*] Corresponding author : olivier.pages@univ-lorraine.fr

zb-Zn$_{1-x}$Be$_x$Se [12] revealed a distinct bimodal Raman signal for the Be-based bond (1-bond→2-mode), beyond the MREI model prediction (1-bond→1-mode) [14]. Within the introduced percolation model (PM) [13] this was explained by sensitivity of bond vibrations to their "*homo*" or "*hetero*" environments, as defined at one-dimension (1D) up to first-neighbors. Accordingly, the O-mode due to a given bond would "feel" fluctuations in the local composition around that bond, inherent to random alloying.

The main features of the PM for Zn$_{1-x}$Be$_x$-chalcogenides can be outlined as follows.

- First – the non-polar O-modes in focus: Out of O-modes, our policy in view to elucidate the phonon mode behavior of a given alloy is to address specifically the non-polar, *i.e.*, purely-mechanical, ones. This is because mechanical oscillators hardly couple and hence preserve the natural diversity of the phonon pattern of a complex system like an alloy.
- Second – a basic 1D scheme: the "*homo*" and "*hetero*" environments are defined at the first-neighbor scale at one dimension (1D) along the linear chain approximation. The justification for a 1D model is that RS/IR-spectroscopies operate at long wavelength [11], therefore only the vibration patterns lattice-translated throughout the crystal can be probed. Such uniformity renders superfluous a realistic description of the crystal at three-dimensional (3D), a mere 1D description suffices.
- Third – a highlight on the small substituent: the bimodal PM-pattern is especially clear for Be-VI (separated by $\delta\omega \sim 6$ meV) and scaled down by one order of magnitude for Zn-VI ($\delta\omega \sim 1.25$ meV). This is because Be-VI involves the small ($R_c$) Be substituent that has more room than Zn to accommodate the bond mismatch in ($\alpha_c, L$) at the local scale, with concomitant impact that the Be-VI phonon frequencies become more diversified.
- Fourth – origin/strength of the PM-doublet: the ordering of the two submodes forming the Be-VI doublet is determined by the local strain. Consider, *e.g.*, $x \sim 1$, for simplicity. An isolated Zn impurity in BeVI creates a local compressive strain because Zn-VI is longer than Be-VI. Hence, the few Be-VI bonds around Zn vibrate in "*hetero*" (ZnVI-like) environment at a higher frequency than the numerous matrix-like Be-VI bonds away from Zn do, vibrating in "*homo*" (BeVI-like) environment. This results in a {dominant-$O_{Be-VI}^{Be}$, minor-$O_{Be-VI}^{Zn}$} PM-doublet for Be-VI at $x \sim 1$, in this order of frequency. The corresponding 1D-oscillators at the first-neighbor scale are $Zn(VI - Be)VI$ and $Be(VI - Be)VI$, respectively, with abundance in the crystal given by $x \cdot (1-x)$ and $x^2$, respectively, assuming a random Be↔Zn substitution. The Raman intensities scale accordingly. Hence, the 1:1, 1:2 and 2:1 intensity ratios between the two Be-VI Raman submodes are achieved at $x \sim 0.5$, $x \sim 0.66$ and $x \sim 0.33$, respectively.

On this basis, the RS spectrum of zb-Zn$_{0.67}$Be$_{0.33}$Se consists of a triple {$O_{Zn-Se}$, minor-$O_{Be-Se}^{Be}$, dominant-$O_{Be-Se}^{Zn}$} signal, the upper two Be-Se modes scaling 1:2 in intensity, as apparent in Fig. S1 (taken from Fig. 2 of Ref. [7]). The inversion between the $R_c$ values of the substituents in Zn$_{1-x}$Be$_x$-chalcogenides and Zn$_{1-x}$Mg$_x$-chalcogenides, with the light Be/Mg substituent of small/large $R_c$ value forming the short/long bond, invites to anticipate an inversion of the RS signal for the nearly composition-matched wurtzite Zn$_{1-x}$Mg$_x$-chalcogenides (main text), supported by *ab initio* calculations ($x \sim 0,1$). In this line, the triple {$O_{Zn-Se}$, minor-$O_{Be-Se}^{Be}$, dominant-$O_{Be-Se}^{Zn}$} signal of Zn$_{0.67}$Be$_{0.33}$Se would transform into an inverted {dominant-$O_{Zn-VI}^{Zn}$, minor-$O_{Zn-VI}^{Mg}$, $O_{Mg-VI}$} RS signal for the nearly composition-matched Zn$_{1-x}$Mg$_x$ chalcogenides, in order of increasing frequency.

## 2. Wurtzite (w) Zn$_{1-x}$Mg$_x$S

Two large Zn$_{1-x}$Mg$_x$S single crystals (photographs are shown in Fig. S2) are studied by INS at low ($x$=0.06) and high ($x$=0.33) alloying. The wurtzite structure is verified by powder X-ray diffraction (Fig. S2). A high structural quality is achieved, even at large Mg incorporation, attested by narrow X-ray diffraction peaks.

### 2.1. *Wurtzite Zn$_{0.67}$Mg$_{0.33}$S – Raman scattering*

The INS study of the w-Zn$_{0.67}$Mg$_{0.33}$S single crystal across the BZ from centre to boundary is based on a preliminary RS study operated near the BZ-centre at room temperature using the 632.8 nm HeNe-laser (Fig. S3). Backward and near-forward RS geometries, probing, respectively the phonon and phonon-polariton (TO) regimes, are realized at normal incidence/detection on (or, across) the ~2mm-thick w-Zn$_{0.67}$Mg$_{0.33}$S slice with parallel faces taken by cleavage from the ingot used for INS, with in-plane $\vec{c}$-axis tilted by ~45° with respect to the ingot axis and out-of-plane $\vec{a}$-axis perpendicular to the cleaved faces.

Specific RS-geometries addressing the phonon symmetries of interest for the PM-test (see main text) are denoted using Porto's notation $\vec{k}_i(\vec{e}_i, \vec{e}_s)\vec{k}_s$ specifying the wavevectors (outside brackets) and polarizations (inside brackets) of the incident laser beam (subscript $i$) and of the scattered light (subscript $s$) [S1].

$B_1^H$ is Raman-silent; the remaining three ($A_1(TO)$, $E_1(TO)$, $E_2^H$) modes are quasi-degenerate within few tenths of meV in both Zn-S and Mg-S spectral ranges. Specific $A_1(TO)$, $E_1(TO)$ and $E_2^H$ RS-insights are achieved by combining the $\bar{X}(Z,Z)X$ forward-RS geometry together with the $X(Z,Z)\bar{X}$ and $X(Z,Y)\bar{X}$ backward-RS geometries, with (X,Y,Z) referring to the ($\vec{a}, \vec{b}, \vec{c}$)-crystal axis. The crossed-polarizations setup is $E_1(TO)$-specific whereas both $A_1(TO)$ and $E_2^H$ are active in parallel polarizations.

By adopting the forward-RS geometry at minimal incidence/scattering angle, $E_2^H$ is cleared from $A_1(TO)$. This is because the transverse-$A_1$ is then probed deep in its polar phonon-polariton-regime ($PP_{A1}$) ultimately close to the BZ-centre, at a much lower frequency than its native non-polar $A_1(TO)$ accessed away from the BZ-centre in a standard backward-RS experiment. As a non-polar mode, $E_2^H$ exhibits no phonon-polariton coupling and hence emerges at the same fixed frequency when probed in either backward-RS or forward-RS. $A_1(TO)$ is eventually inferred by difference between the $\bar{X}(Z,Z)X$ and $X(Z,Z)\bar{X}$ Raman signals, which probe the $PP_{A1}$ and $A_1(TO)$ variants of the transverse-$A_1$, respectively. The $A_1(TO) - E_1(TO)$ frequency gap is nearly the same, *i.e.*, ~1.75 meV, for Zn-S and Mg-S, whereby $E_2^H$ emerges near $E_1(TO)$, within few tenths of meV, either below (Zn-S) or above (Mg-S) it.

In Fig. S3, $A_1(TO)$, $E_1(TO)$ and $E_2^H$ uniformly exhibit a crude MREI (1-bond→1-mode) signal in each (Zn-S, Mg-S) spectral range. Hence, the PM-test on w-Zn$_{0.67}$Mg$_{0.33}$S is negative by RS, in either phonon symmetry. However, the Zn-S Raman signal – of main interest for the PM-test (see main text) – is obscured by a pronounced Fano interference manifested by a deep antiresonance (marked by an asterisk) situated in between the Zn-S and Mg-S signals, motivating further INS insight.

The Fano antiresonance is released in the INS spectrum of Zn$_{0.67}$Mg$_{0.33}$S taken near the BZ-centre at 10K in $E_1(TO)$ (Fig. 2) and $A_1(TO)$ (Fig. S8) symmetries. As low temperatures dramatically impact the two-phonon scattering, but not so much the one-phonon scattering, the Fano release at 10K presumably marks the involvement of a two-phonon continuum in the Fano process.

### 2.2. *Wurtzite Zn$_{1-x}$Mg$_x$S – Inelastic neutron scattering and ab initio calculations*

### 2.2.a. *Wurtzite Zn$_{1-x}$Mg$_x$S – Ab initio* insight at $x\sim 1$

The *ab initio* test done in the main manuscript on the Zn-S signal in the Mg-dilute limit of Zn$_{1-x}$Mg$_x$S ($x\sim 0$, Fig. S5) is presently transposed to the Mg-S signal in the Zn-dilute limit ($x\sim 1$, Fig. S6), to explore a possible PM-type fine structuring of the Mg-S O-mode in a simple case. The advantage of placing the *ab initio* study at the impurity limit is that a given mode can be assigned directly from its wavevector. Away from the impurity limit, this is not possible because the wavevectors are blurred by the alloy disorder.

The discussion done at $x\sim 0$ (main text) can be directly replicated at $x\sim 1$, only that the local strain then becomes inverted, *i.e.*, tensile in character (inset of Fig. S6a). This results in an inverted {dominant-$L_{Mg-S}^{Mg}$, minor-$L_{Mg-S}^{Zn}$} doublet (1-bond→2-$L$) for the Mg-S bond length ($L$) at $x\sim 1$, leading to an inverted {minor-$O_{Mg-S}^{Zn}$, dominant-$O_{Mg-S}^{Mg}$} PM-duo for the Mg-S O-mode (1-bond→2-mode), in order of increasing length/frequency (Figs. S6a and 6b). The atom displacement behind the $A_1$ minor-$O_{Mg-S}^{Zn}$ mode (a snapshot is shown in Fig. S6c) selectively involves an effective bond stretching (O-mode) along $\vec{c}$ of the few Zn-S bonds forming the first-neighbor shell of the Mg impurity. This proves the highly-localized character of this mode.

Not surprisingly (see main text), the phonon splitting is smaller (by roughly a factor two) for Mg-S ($\delta\omega_{Mg-S}\sim 2.83$ meV, $x\sim 1$ – Fig. S6) than for Zn-S ($\delta\omega_{Zn-S}\sim 4.10$ meV, $x\sim 0$ – Fig. S5), due to a larger covalent radius for Mg as compared to Zn. Remarkably, the Mg-S duo at $x\sim 1$ is better resolved when polarized transverse to $\vec{c}$ ($\delta\omega_{Mg-S}^{\perp\vec{c}}\sim 2.83$ meV) than along $\vec{c}$ ($\delta\omega_{Mg-S}^{\|\vec{c}}\sim 1.51$ meV). Such polarization effect is absent for the Zn-S duo probed at $x\sim 0$.

Altogether, the *ab initio* insights at $x\sim 0$ (main text) and $x\sim 1$ (this Sec.) support PM-duos for Zn-S and Mg-S in w-Zn$_{1-x}$Mg$_x$S. However, as $\delta\omega_{Mg-S} < \delta\omega_{Zn-S}$ the PM-splitting should show up clearly especially for Zn-S in the INS spectra of w-Zn$_{0.67}$Mg$_{0.33}$S, and not so much for Mg-S. In fact, $\delta\omega_{Zn-S}$ is equally large in all phonon symmetries found relevant for the PM-test (Fig. S5b), exceeding the natural width at half maximum of the individual phonon peaks, so that the same distinct Zn-S PM-duo is expected to show up in all tested symmetries in experiment. The situation is not so favorable for Mg-S. As $\delta\omega_{Mg-S}^{\parallel \vec{c}} < \delta\omega_{Mg-S}^{\perp \vec{c}}$, the best chance to detect experimentally the PM-duo for Mg-S is via $(E_1(TO), E_2^H)$, and not so much via $(A_1(TO), B_1^H)$.

*2.2.b. Wurtzite $Zn_{1-x}Mg_xS$ – O-dispersions*

Overviews of the w-$Zn_{1-x}Mg_xS$ phonon dispersions related to the non-polar $E_2^H, A_1(TO)$ and $B_1^H$ O-modes across the BZ are regrouped in triptychs (adopting the format used in Fig. 2) in Figs. S7-to-S9. The used format emphasizes a comparison between measured (INS) and calculated (*ab initio* – the used special quasirandom structures are shown in Fig. S4) dispersions at high ($x$=0.33) and low ($x$=0.06) alloying, completing the $E_1(TO)$ insight provided in Fig. 2. The pure $A_1(TO)$, $E_1(TO)$ and $E_2^H$ symmetries are addressed in experiment, so that the reported *ab initio* dispersions are also pure in the triptychs, for a direct comparison. In experiment, the $B_1^H$ dispersion along $\overrightarrow{c^*}$ can be confused by (equally permitted) $A_1(LO)$ (Fig. S9a). A first level of clarification is achieved *ab initio* by calculating the $B_1^H + A_1(LO)$ $q$-projected phonon density of states ($q$-PhDOS) per Mg cation, emphasizing the Mg-S signal (Fig. S9c), to compare directly with experiment. More clarification is needed in the sensitive Zn-S spectral range where the PM-duo is most likely to occur – by analogy with the $A_1(TO)$, $E_1(TO)$ and $E_2^H$ symmetries. This is achieved by calculating *ab initio* the pure-$B_1^H$ $q$-projected phonon density of states per Zn cation (Fig. S9b). Hence, in total, the INS vs. *ab initio* insight for the $B_1^H + A_1(LO)$ symmetries is enlarged to a quartic involving two *ab initio* panels (Fig. S9).

First, we discuss the $E_2^H, A_1(TO)$ and $E_1(TO)$ INS vs. *ab initio* triptychs.

At the BZ-centre, the INS vs. *ab initio* comparison in the last two symmetries is critical in various respects. Before the discussion starts, we recall a basic selection rule that only the modes polarized along the transferred wavevector $\vec{Q} = \vec{G} + \vec{q}$ are detected by INS, using the notation introduced in the main text. By adopting a neutron scattering geometry likely to address the TO modes near the BZ-centre $\Gamma$ ($q$~0), the $A_1(LO)$ and $E_1(LO)$ modes transiently show up strictly at $\Gamma$ ($q$=0). This is because at this limit ($q = 0$) the distinction between the transverse (perpendicular to $\vec{q}$) and longitudinal (parallel to $\vec{q}$) vibrations disappears [11]. As soon as departing from $\Gamma$, $A_1(LO)$ and $E_1(LO)$ immediately vanish, so that only the genuine transverse $A_1(TO)$ and $E_1(TO)$ survive. We emphasize that the TO modes in question are non polar, *i.e.*, purely mechanical in character. The polar TO modes, *i.e.*, the phonon-photon coupled modes known as phonon-polaritons, cannot be detected by INS. Indeed, by force of the quasi vertical dispersion of a photon, the phonon-polaritons are confined extremely close to the $\Gamma$ ($q = 0$) point of the first BZ ($G = 0$), *i.e.*, within the first few one per ten thousand ($1^0/_{000}$) of the BZ size [11]. Generally, the INS experiments are operated in special BZ remote from the first BZ where the neutron scattering cross section is maximized and/or best resolved for a given mode. No chance to detect phonon-polaritons is given there. Even by placing the INS study near the $\Gamma$ ($q$~0) point of the first BZ ($G = 0$), the phonon-polariton would remain out of reach by INS. It is not a mere technical problem of a limited $\vec{q}$-resolution by INS. The limitation is inherent to INS, due to the cited INS selection rule above that the probed vibrations by INS are polarized along $\vec{G} + \vec{q}$ [12]. By force of the latter rule, near the $\Gamma$ ($q$~0) point of the first BZ ($G = 0$) where the photon-polariton coupling develops, the only vibrations that can be probed by INS are polarized along $\vec{G} + \vec{q}$, reducing to $\vec{q}$ in this case, hence longitudinal in character and not transverse. In brief, an INS experiment done near the $\Gamma$ ($q$~0) point of the first BZ ($G = 0$) would probe the LO modes only, not the phonon-polaritons. Note that neither the phonon-polaritons nor the LO modes can be addressed *ab initio* near the centre ($q$~0) of any BZ ($G = 0$ or $\neq 0$). This is because in its current version the used SIESTA code fails to handle the macroscopic ($q$~0) transverse (phonon-polariton) or longitudinal (LO) electric fields due to the ionicity of the chemical bonding in a wurtzite crystal [11]. Immediately out of the BZ-centre, the phonon-polaritons disappear but the electric field carried by the polar $A_1(LO)$ and $E_1(LO)$ modes is restored *ab initio*. An explicit example is given in Fig. S9c.

Precisely, by departing from the BZ-centre, the INS signal of w-$Zn_{0.67}Mg_{0.33}S$ clarifies, showing a distinct PM-duo for Zn-S in the pure $E_1(TO)$ (Fig. 2), $E_2^H$ (Fig. S7) and $A_1(TO)$ (Fig. S8) symmetries, that gains in resolution by approaching the BZ-boundary, as earlier observed by INS for the Be-Se PM-duo of $Zn_{0.67}Be_{0.33}Se$ [7]. A careful examination of the w-$Zn_{0.67}Mg_{0.33}S$ INS spectra even reveals a compact but distinct {dominant-$O_{Mg-S}^{Zn}$, minor-$O_{Mg-S}^{Mg}$} PM-duo for Mg-S near the BZ-boundary in symmetries $E_2^H$ and $E_1(TO)$, hardly resolved in the $A_1(TO)$ symmetry. This ideally conforms to expectations – referring to the *ab initio* insight at $x$~1 (Fig. S6), only that the $O_{Mg-S}^{Zn}$ vs. $O_{Mg-S}^{Mg}$ intensity balance is opposite at $x$=0.33 (Fig. 2) and $x$~1 (Fig. S6). This is because the Zn-like environment is sub-represented at $x$~1 and over-represented at $x$=0.33 – and *vice versa* for the Mg-like environment, with concomitant impact on the INS/*ab initio* intensities.

We turn to the [$B_1^H + A_1(LO)$] INS vs. *ab initio* quartic (Fig. S9). In experiment the situation for $B_1^H$ is obscured by $A_1(LO)$, as already mentioned. The main trend is that on departing from the BZ-centre, $B_1^H$ and $A_1(LO)$ progressively converge in each spectral range, until the perfect degeneracy is eventually achieved at the BZ-boundary, as observed with w-ZnS [35] and w-MgS [36,37]. The convergence is especially clear for Mg-S, both in experiment (Fig. S9a) and *ab initio* (Fig. S9c). Interestingly, halfway the Brillouin zone, the so far degenerate Zn-S $A_1(LO)$ and Mg-S $B_1^H$ modes break away in converging towards the lower Zn-S $B_1^H$ and upper Mg-S $A_1(LO)$ modes, respectively. This clarifies the assignment of modes. A very broad ZnS-like $B_1^H$ feature, asymmetrical on its high-frequency side (Fig. S9a), is revealed, suggesting a {dominant-$O_{Zn-S}^{Zn}$, minor-$O_{Zn-S}^{Mg}$} $B_1^H$ PM-duo for Zn-S. While helpful, the INS insight is not decisive due to the spurious $A_1(LO)$ signal, motivating to search for a direct *ab initio* insight into the pure $B_1^H$ signal of Zn-S (Fig. 9b). The *ab initio* Zn-S signal is unimodal at the BZ-centre, but progressively develops into a distinct PM-duo at the BZ-boundary, similar in every respect to that observed by INS and *ab initio* in the $E_2^H, A_1(TO)$ and $E_1(TO)$ symmetries. Hence, experiment (INS) and *ab initio* used in combination prove the existence of the Zn-S PM-duo in $B_1^H$ symmetry on approach to the BZ-boundary, further supporting the view that PM-duos are better resolved at this limit. No distinct PM fine structure is evidenced by performing the same pure-$B_1^H$ *ab initio* calculations for the Mg cation (not shown). This is not surprising since $\delta\omega_{Mg-S}$ is so small in $B_1^H$ symmetry, as attested *ab initio* ($x$~1).

In brief, the same distinct Zn-S {dominant-$O_{Zn-S}^{Zn}$, minor-$O_{Zn-S}^{Mg}$} PM-duo shows up in all relevant non-polar $A_1(TO), E_1(TO), E_2^H$ and $B_1^H$ symmetries (corresponding to a large $\delta\omega_{Zn-S}$ value), in experiment and *ab initio*. A more compact Mg-S {dominant-$O_{Mg-S}^{Mg}$, minor-$O_{Mg-S}^{Zn}$} PM-duo is likewise transiently observed in experiment by INS near the BZ-boundary in the most relevant ($E_1(TO), E_2^H$) symmetries (corresponding to the maximal $\delta\omega_{Mg-S}$ value, as anticipated *ab initio* at $x$~1). The PM-test on the O-modes of w-Zn$_{0.67}$Mg$_{0.33}$S is thus decisive in the positive sense.

Altogether, the joined observations of the Zn-S (Fig. 2) and Mg-S PM-duos by INS in w-Zn$_{0.67}$Mg$_{0.33}$S (Fig. S7) together with the observation of similar, yet inverted, PM-duos in the RS spectra of the composition-matched zb-Zn$_{0.67}$Be$_{0.33}$Se alloy (Fig. S1) demonstrates the generic character of the PM, being neither bond- nor alloy-dependent.

### 2.2.c. Wurtzite Zn$_{1-x}$Mg$_x$S – A/O-dispersions

The phonon dispersions of the bond-collective (2-bond→1-modes) modes of w-Zn$_{0.67}$Mg$_{0.33}$S (covering the actual acoustic modes plus the $E_2^L$ and $B_1^L$ O-modes) measured by INS (upper panels) and calculated *ab initio* (lower panels) along the $\vec{a}$ and $\vec{c}$ crystal axis are displayed in Figs. S10 and S11, respectively. Despite such modes are not relevant to test the PM because they are bond-insensitive, and thus *a fortiori* "blind" to the local environment of a bond, we cannot escape a brief discussion of their dispersions, for the sake of completeness.

#### 2.2.c.1 A/O-dispersions near the BZ-boundary

Remarkably, on approaching the BZ-boundary, the TA and LA modes exhibit opposite trends along $\vec{c^*}$ (Fig. S10) and $\vec{a^*}$ (Fig. S11) *i.e.*, they improve/sharpen (dashed circles in Fig. 1) and degrade/broaden (hollow circles in Fig. 1), respectively. Both trends are attested in experiment and *ab initio*, hence intrinsic to random Zn$_{1-x}$Mg$_x$S-alloying. For example, in experiment $A_1(LA)$ and $E_1(TA)$ sharpen on approaching the BZ-boundary along $\vec{c^*}$ (Fig. S10a and S10b). The trend is also attested *ab initio* (Fig. S10c and S10d). Alternatively, the degradation at the BZ-boundary along $\vec{a^*}$ occurs for $A_1(TA)$, in experiment (Fig. S11b) and *ab initio* (Fig. S11d). Hence, the genuine acoustic modes of hexagonal w-Zn$_{0.67}$Mg$_{0.33}$S do not become fatally overdamped at the BZ-boundary, at variance with observations on cubic HEA [4-6] and cubic zb-Zn$_{0.67}$Be$_{0.33}$Se alloy [7]. When they mix with the bond-collective O-modes of same polarization, such as $E_1(TA)$ with $E_2^L$ polarized perpendicular to $\vec{c^*}$ and $A_1(LA)$ with $B_1^L$ polarized along $\vec{c^*}$, the natural tendency of the O-modes to sharpen near the BZ-boundary – as observed with the PM-duos of zb-Zn$_{0.67}$Be$_{0.33}$Se [7] and w-Zn$_{0.67}$Mg$_{0.33}$S (this work) – takes over.

#### 2.2.c.2 A/O-dispersions halfway the BZ

Similar discrepancy occurs halfway the BZ. For example, along $\vec{c^*}$, $A_1(LA)$ (Fig. S10, left panels) and $E_2^L$ (Fig. S10, right panels) do not "see" each other on crossing, neither in experiment (Fig. S10, top panels) nor *ab initio* (Fig. S10, bottom panels), due to their polarizations being along $\vec{c^*}$ and perpendicular to $\vec{c^*}$, respectively. Hence, the acoustic modes of w-Zn$_{0.67}$Mg$_{0.33}$S do not fatally degrade on crossing with O-modes halfway their dispersion, at variance with recent observations on a disordered cubic ceramics [38]. We refer to the dramatic overdamping of the $\Sigma(TA_2)$ mode (Fig. 4 of Ref. [38]), "stopped" halfway in its dispersion on progressing towards a defect-induced O-mode (Fig. 5 of Ref. [38]).

In contrast, being both polarized perpendicular to $\vec{c^*}$, $E_1(LA)$ and $E_2^L$ do effectively couple/interfere on crossing each other halfway their dispersion along $\vec{a^*}$. This gives rise to two coupled modes in mutual repulsion, from now on labelled ($-$) and ($+$) in order of increasing frequency. In fact, the INS data acquired at low frequency in the longitudinal symmetry along $\vec{a^*}$ (the scattering geometry is sketched out, for clarity), addressing both the longitudinal-optical $E_2^L$ and longitudinal-acoustic $E_1(LA)$ modes, reveal two separate (longitudinal) modes that regularly shift apart on departing from the BZ-centre and progressing towards the BZ-boundary Fig. S11a). The trend is independently confirmed *ab initio* data of cumulated $E_2^L + E_1(LA)$ $\vec{q}$-projected PhDOS per anion along $\vec{a^*}$ (Fig. S11c). In brief, no crossing of the $E_1(LA)$ and $E_2^L$ modes is evidenced, neither in experiment nor *ab initio*, suggesting instead an anticrossing due to an effective coupling.

In this case, each individual ($\pm$) coupled mode should jointly manifest the $E_2^L$ and $E_1(LA)$ characters across the $q$-domain spanning the coupling regime. This cannot be checked experimentally, but an *ab initio* insight can be achieved by addressing separately the $\vec{q}$-projected $E_2^L$ (dotted curves) and $E_1(LA)$ (solid curves) PhDOS per cation (summed across Zn and Mg – completing the insight per anion given above) along $\vec{a^*}$ (Fig. S12a). Strictly at $q$=0, the $E_1(LA)$ and $E_2^L$ modes are decoupled, emerging as distinct features at zero frequency and at ~10 meV, respectively. Once departing from the BZ-centre (within the set *ab initio* $q$-resolution), each of the two apparent *ab initio* modes takes on a $E_1(LA) - E_2^L$ mixed character. The amount of mixing eventually achieves maximum around the BZ-boundary (~50%).

A crude description of the INS and *ab initio* coupled ($\pm$) modes in their $q$-dependence (Fig. S12a), covering both (i) the frequencies $\omega_\pm$ and (ii) the amount of $E_1(LA) - E_2^L$ mixing per coupled mode across the BZ (Fig. S12c), is phenomenologically achieved within a simple model of two coupled harmonic oscillators (mass + restoring force), *i.e.*, one per symmetry, to fix ideas. The modeling is mostly concerned with the *ab initio* data, for which both informations (i) (hollow symbols – Fig. S12b, marred by error bars) and (ii) (in relation to Fig. S12c) are available, whereas only information (i) is available from the raw INS data (solid symbols, Fig. S12b).

At least $E_2^L$ is non polar (the net dipole generated by $E_2^L$ per primitive cell is strictly zero), *i.e.*, purely mechanical in character. Hence, the coupling has to be mechanical in character. $E_1(LA)$ and $E_2^L$ are bond-collective, in that they both involve the dynamics of the two bond species (Mg-S) and (Zn-S) of Zn$_{0.67}$Mg$_{0.33}$S at the same frequency. It can be considered that they both refer to the vibration of the same virtual (Zn$_{1-x}$Mg$_x$)-S bond with physical properties averaged over the actual Zn-S and Mg-S bonds depending on their fractions in the crystal, in the spirit of the virtual crystal approximation. The reduced mass of such virtual bond involved as well in the $E_1(LA)$ and $E_2^L$ modes, is denoted $\mu$ hereafter. The equation of motion per oscillator $i$ in presence of oscillator $j$ then takes the generic form $\mu \ddot{u}_i = -k_i u_i - k'(u_i - u_j)$, including a mechanical restoring force characteristic of the raw uncoupled oscillator $i$ (first term) plus a mechanical coupling with the neighboring oscillator $j$ (second term). The dynamical matrix expresses as

$$\widetilde{M} = \begin{pmatrix} \omega_1^2 + \omega'^2 - \omega^2 & -\omega'^2 \\ -\omega'^2 & \omega_2^2 + \omega'^2 - \omega^2 \end{pmatrix}, \quad (1)$$

where $\omega_i^2 = \mu^{-1} k_i$ represents the square frequency of the uncoupled oscillator-$i$ and $\omega'^2 = \mu^{-1} k'$ is the square frequency characteristic of the mechanical coupling. The $E_2^L$ frequency ($\omega_2$) is taken constant in first approximation, fixed to the *ab initio* value at the BZ-centre (dotted curve). Both the $E_1(LA)$ frequency ($\omega_1$) and $\omega'$ are zero at $q$=0. This is because as an acoustic mode $E_1(LA)$ does not propagate at $q$=0, and $E_1(LA)$ and $E_2^L$ are uncoupled at $q$=0 (see above), respectively. Linear $q$-dependencies are assumed for $\omega_1$ and $\omega'$, in a crude approximation, by analogy with $E_1(LA)$, and because the $E_1(LA)-E_2^L$ coupling reaches maximum at the BZ-boundary (see above), respectively. The slopes are the only adjustable parameters, governing (i) $\omega_\pm$ in their $q$-dependence and (ii) how much (in percent) $E_1(LA)$ and $E_2^L$ are involved into each ($\pm$) coupled mode at a given $q$ value.

Informations (i) and (ii) are accessible via the eigenvalues and the eigenvectors of $\widetilde{M}$, respectively, as detailed in Ref. [S2] Basically, by writing the off-diagonal and diagonal terms of the dynamical matrix as $E_{i,j} - \omega^2$ and $V$ ($i,j$=1,2), respectively, the frequencies of the coupled modes express as $\omega_\pm^2 = E \pm \sqrt{\Delta^2 + V^2}$, depending on the proximity with the resonance $\Delta = (E_1 - E_2)/2$ and on the strength of coupling $V$, with $E = (E_1 + E_2)/2$. The unit $|u_\pm\rangle$ wavevectors are orthogonal, following from, *e.g.*, $|Q_+\rangle = \begin{pmatrix} \cos\theta \\ \sin\theta \end{pmatrix}$, with $\cos^2\theta$ and $\sin^2\theta$ indicating how much, in percent, the raw-uncoupled $|u_1\rangle$ and $|u_2\rangle$ oscillators are involved in $|u_+\rangle$, respectively, and *vice versa* for $|u_-\rangle$. An explicit form of $\cos\theta$ depending on $V$ and $\Delta$ is given in Ref. [S2] (refer to Eq. 5 therein). A fair description of the

coupled modes ($\pm$) in mutual repulsion (Fig. S12b – plain curves) is achieved by considering that both $\omega_1$ and $\omega'$ scale as $\omega = 40 \times q$. Despite its simplicity, the as-set model fairly replicates in main lines the *ab initio* (full symbols – taken from Fig. S12a) and experimental (hollow symbols – taken from Fig. S11a) $q$-dependencies of $\omega_-$ and $\omega_+$, and also the large amount of $E_1(LA) - E_2^L$ mixing observed *ab initio* (~50%, based on quasi similar intensities of the plain and dotted signals in Fig. S12a) within each coupled mode ($\pm$) across the BZ.

**Supplementary material-only references**

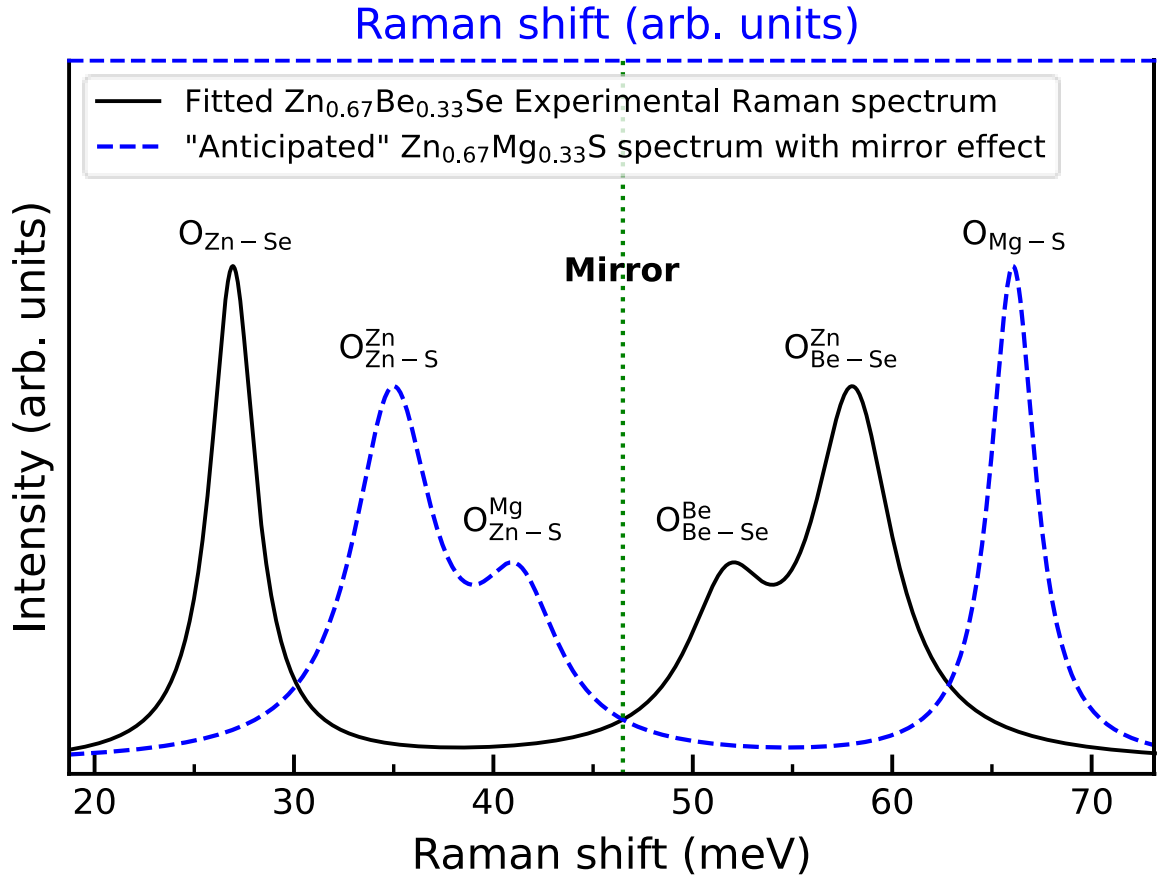

FIG. S1. Reference Raman signal of zincblende $Zn_{0.67}Be_{0.33}Se$. Reference Raman insights into the non-polar O-modes (TO) of zincblende (zb) $Zn_{0.67}Be_{0.33}Se$ and wurtzite (w) $Zn_{0.67}Mg_{0.33}S$. The triple $\{TO_{Zn-Se}, TO^{Zn}_{Be-Se}, TO^{Be}_{Be-Se}\}$ signal (dark-plain curve) including the Be-Se PM-duo, fitted from the experimental zb-$Zn_{0.67}Be_{0.33}Se$ RS signal reported in Fig. 2 of Ref. [7], is labeled by specifying the bond vibration (subscript) and the local bond environment (superscript). The light Be and Mg substituents oppositely exhibit small and large $R_c$ values forming short and long bonds, respectively, so that the w-$Zn_{0.67}Mg_{0.33}S$ Raman signal (dashed-blue curve) can be anticipated as a mirror image of the zb-$Zn_{0.67}Be_{0.33}Se$ Raman signal, in fact derived by merely reversing the abscissa axis, as schematically shown. The horizontal axis of the Raman-image (w-$Zn_{0.67}Mg_{0.33}S$) is not labeled to emphasize that the parallel between both spectra is only phenomenological.

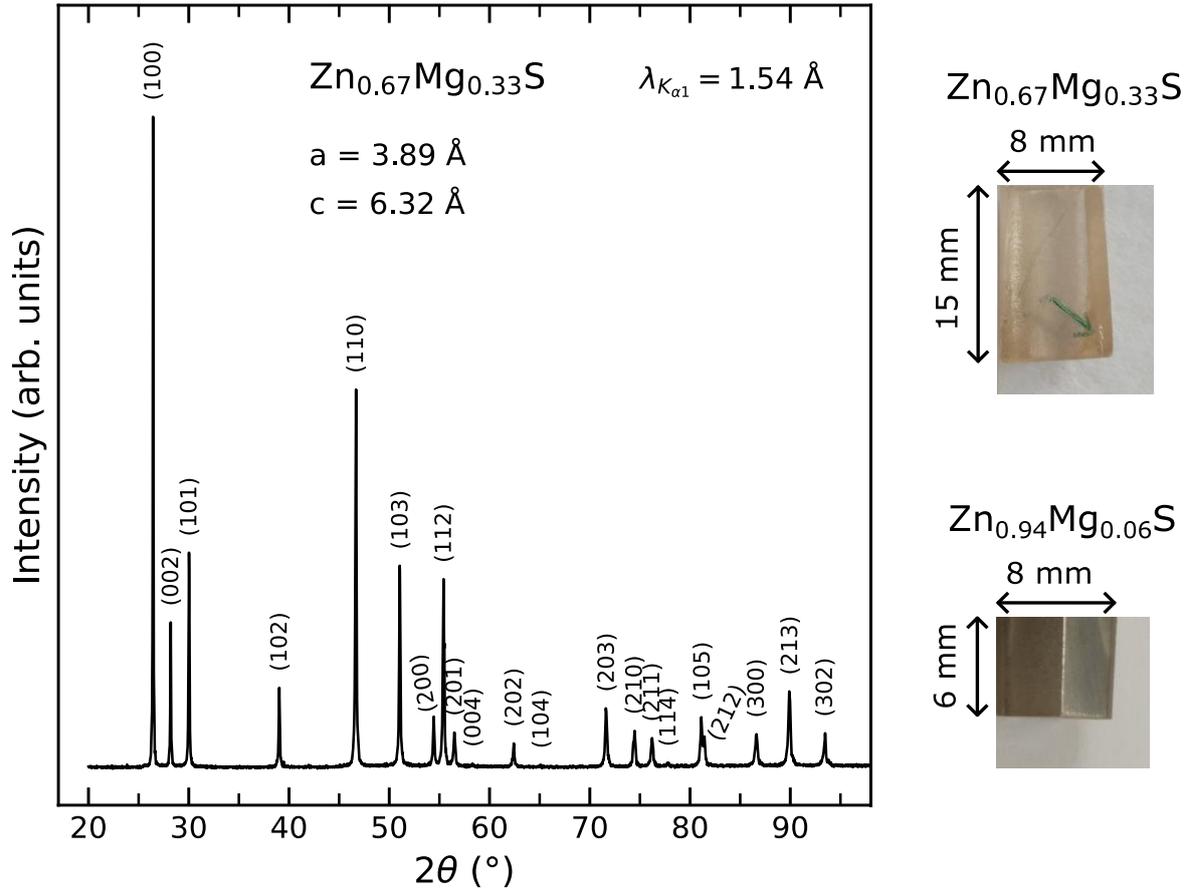

FIG. S2. $Zn_{1-x}Mg_xS$ crystals – size and structure. Powder X-ray diffractogram obtained at ambient conditions with w-$Zn_{0.67}Mg_{0.33}S$. The individual peaks are labelled via the (hkl) Miller indices. The corresponding lattice parameters (a and c) are indicated. Photographs of the cleaved semi-cylindrical $Zn_{1-x}Mg_xS$ single crystals studied by INS are shown on the right. The $\vec{c}$ axis of w-$Zn_{0.67}Mg_{0.33}S$, co-oriented by conoscopy, X-ray diffraction and neutron diffraction *in situ*, lies in-plane, *i.e.*, within the cleaved face, tilted by ~45° with respect to the ingot axis, as indicated by the arrow.

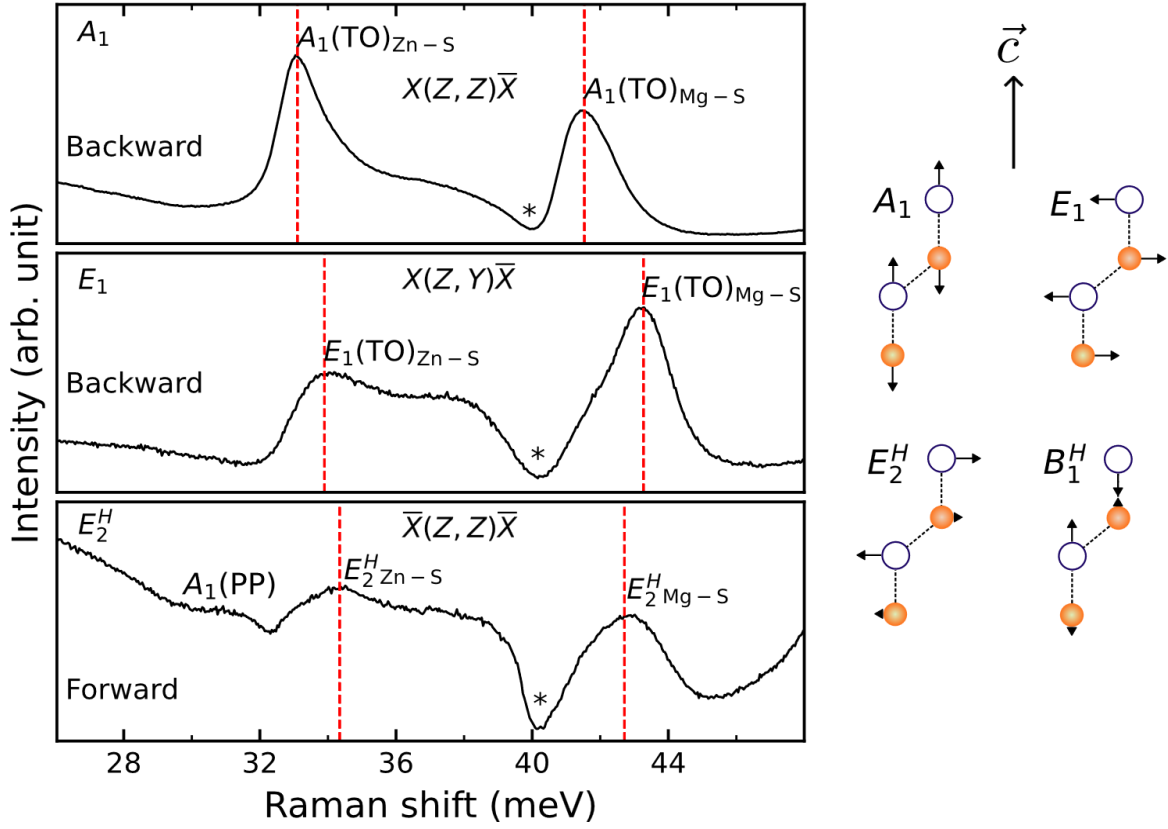

FIG. S3. Polarized Raman spectra of wurtzite (w) Zn$_{0.67}$Mg$_{0.33}$S. Polarized Raman spectra addressing specific non-polar Raman active $E_2^H$, $A_1(TO)$ and $E_1(TO)$ O-modes of w-Zn$_{0.67}$Mg$_{0.33}$S taken at normal incidence/detection onto/through parallel cleaved crystal faces in backward or forward scattering geometries. $A_1(TO)$ and $E_2^H$ are allowed and $E_1(TO)$ is forbidden in the $X(Z,Z)\bar{X}$ and $\bar{X}(Z,Z)\bar{X}$ geometries. $A_1(TO)$ and $E_2^H$ are forbidden and $E_1(TO)$ is allowed in the $X(Z,Y)\bar{Z}$ geometry. In forward scattering the polar Zn-S and Mg-S $A_1(TO)$ soften on entering the phonon-polariton (PP) regime, revealing the non-polar $E_2^H$ (not subject to phonon-polariton coupling). In each symmetry, the RS signal is obscured by a sharp antiresonance (marked by a star) due to a Fano interference. For each phonon symmetry, the atom vibration pattern with respect to the singular $\vec{c}$ crystal axis is sketched out.

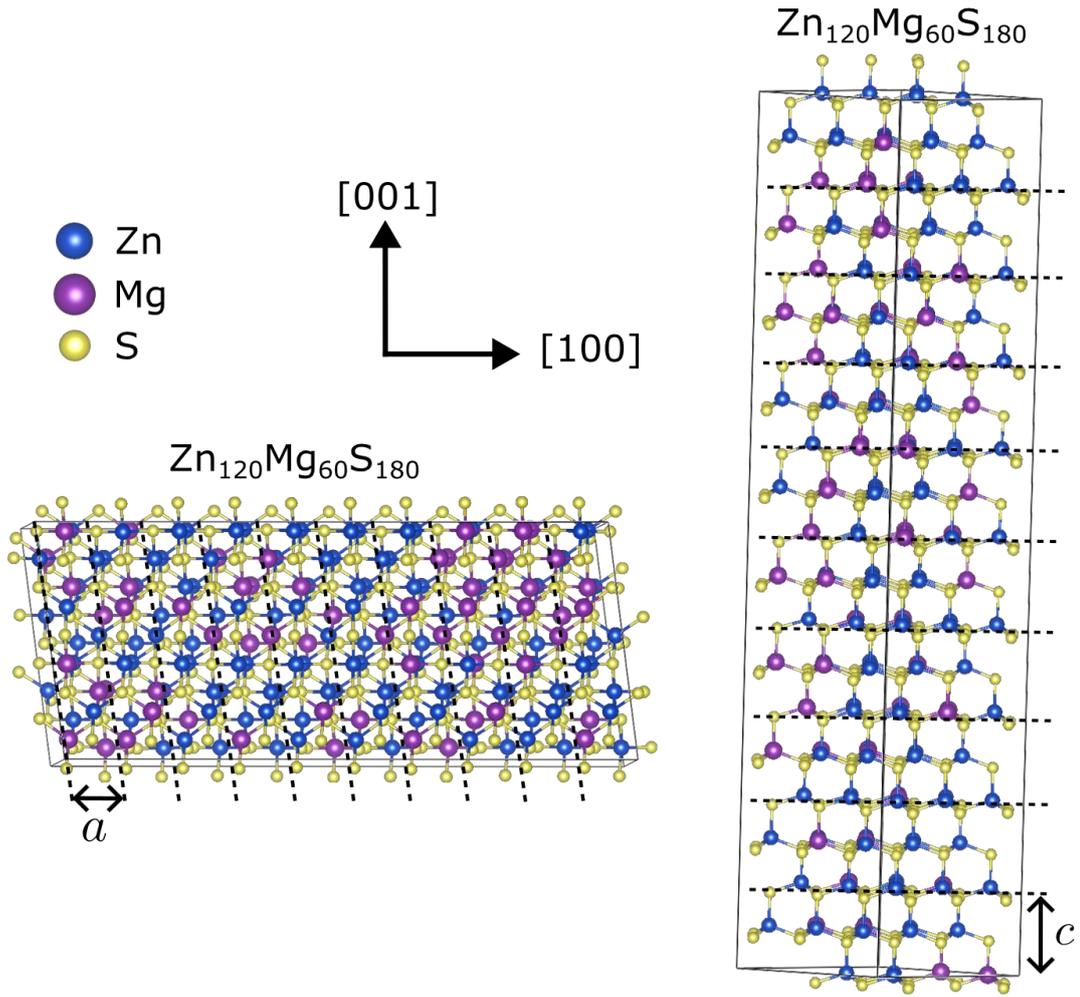

FIG. S4. Wurtite $Zn_{0.67}Mg_{0.33}S$ special quasirandom structures. Special quasirandom structures optimized to a random Mg↔Zn (33 at.% Mg) substitution (using ATAT) elongated along the [100] and [001] directions of the real space in 10 planar sequences (marked by dotted lines) of 9 primitive cells arranged in a square-like pattern, used to calculate the $q$-projected phonon density of states ($q$-PhDOS) along the $\vec{a^*}$ and $\vec{c^*}$ directions of reciprocal space, respectively.

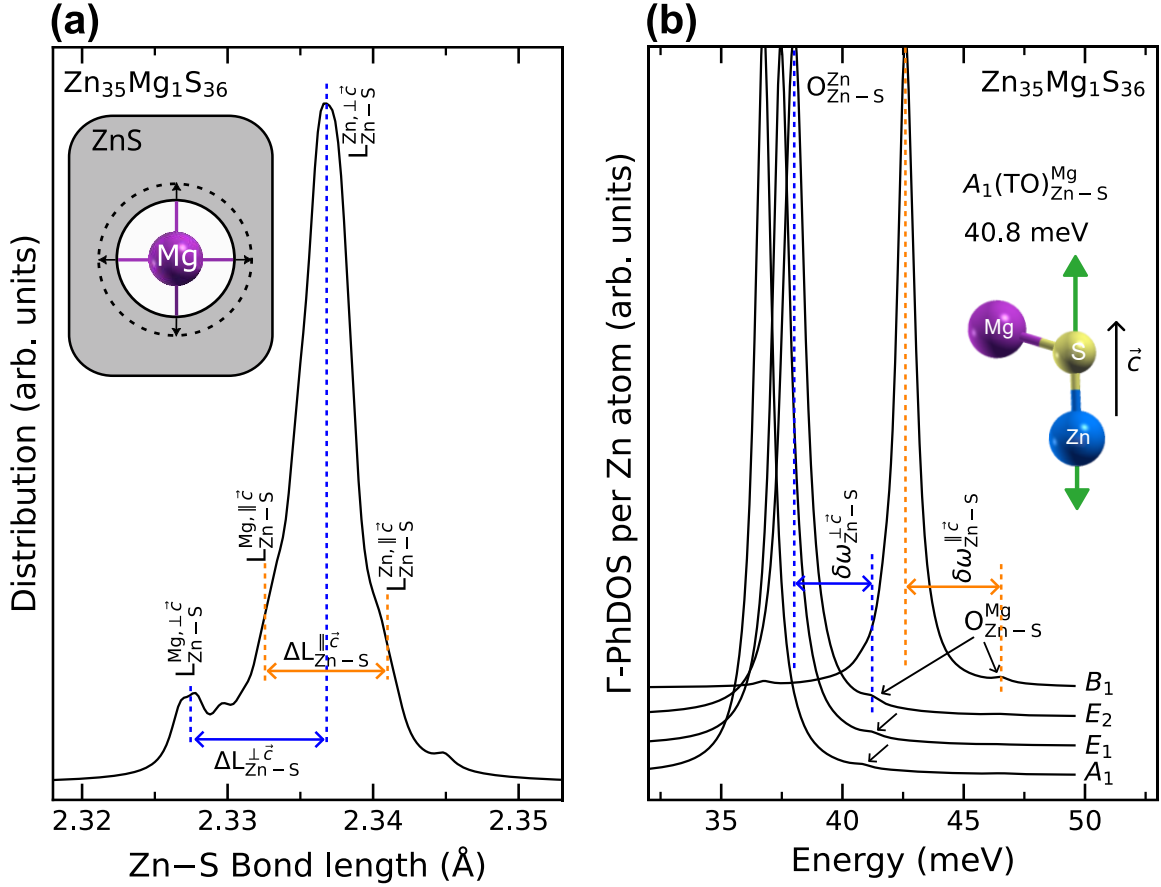

FIG. S5. *Ab initio* insight into the origin of the Zn-S PM-duo of wurtzite $Zn_{1-x}Mg_xS$ ($x\sim0$). (a) *Ab initio* distribution of Zn-S (subscript) bond lengths ($L$) in a large ZnS-like supercell involving a unique Zn↔Mg substitution corresponding to different bond lengths near Mg (*hetero* environment) and far from it (*homo* environment, first superscript) either along the $\vec{c}$ axis or perpendicular to it (second superscript). Inset: The long Mg-bonds generate a compressive strain within their first-neighbour Mg-shell, delimited by a dotted curve, as sketched out (omitting the anisotropy of the crystal structure). (b) Corresponding PM-duo of Zn-S optical ($O$) mode, labelled using the same subscript/superscript code, generated *ab initio* by projecting the phonon density of states per Zn atom at the Brillouin zoneC (Γ-PhDOS). The $A_1(TO)$, $E_1(TO)$, $B_1^H$ and $E_2^H$ symmetries are separately addressed, as indicated. Inset: The local $A_1$-like Zn-S stretching in presence of Mg is sketched out to fix ideas (the exact atom displacement looks like that shown in Fig. S6c except that the Mg and Zn roles should be reversed).

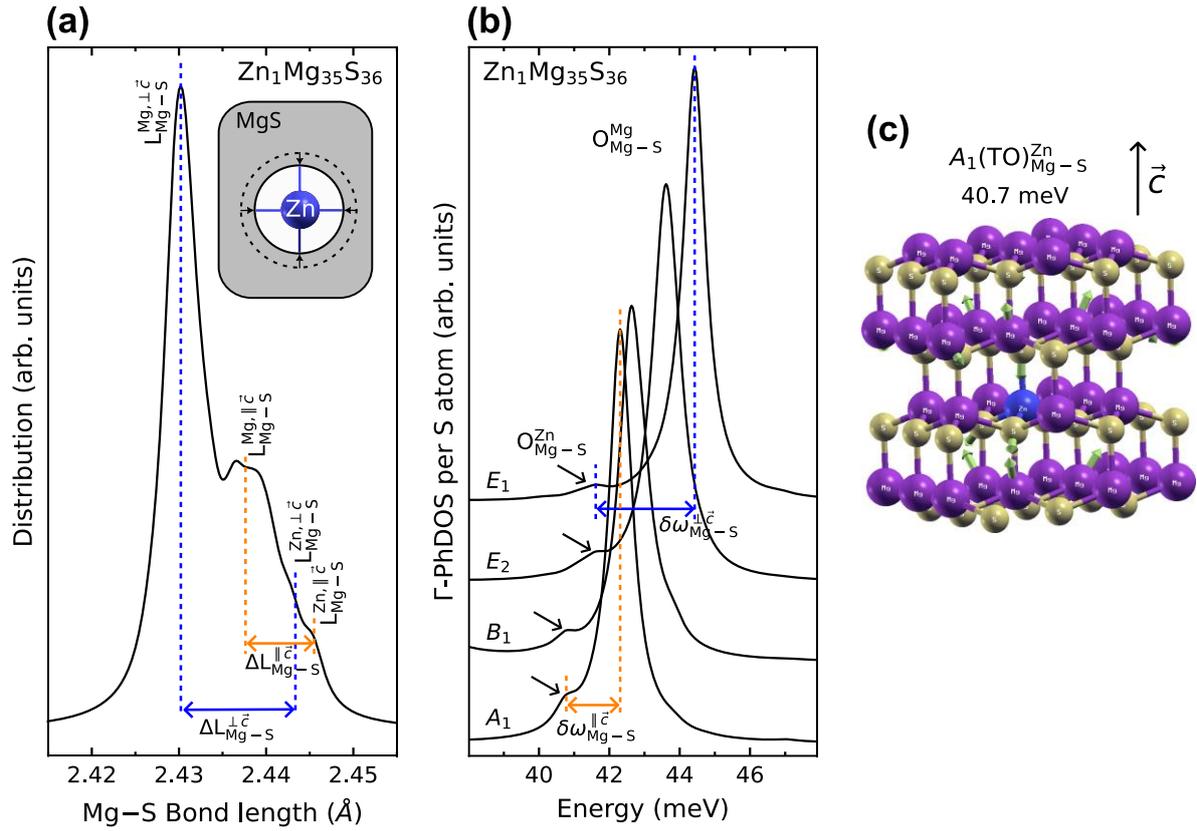

FIG. S6. *Ab initio* insight into the origin of the Mg-S PM-duo of wurtzite $Zn_{1-x}Mg_xS$ ($x\sim1$). (a) *Ab initio* distribution of Mg-S bond lengths ($L$) in a large MgS-like supercell involving a unique Zn ↔Mg substitution, corresponding to different bond lengths near Zn (*hetero* environment) and far from it (*homo* environment, first superscript) either along the $\vec{c^*}$ axis or perpendicular to it (second superscript). Inset: The short Zn-bonds generate a tensile strain within their first-neighbour Zn-shell, delimited by a dotted curve, as sketched out (omitting the anisotropy of the crystal structure). (b) Corresponding PM-duo of Mg-S optical ($O$) mode, labelled by using the same subscript/superscript code, generated *ab initio* by projecting the phonon density of states per S atom at the Brillouin zone centre (Γ-PhDOS). The $A_1(TO)$, $E_1(TO)$, $B_1^H$ and $E_2^H$ symmetries are separately addressed, as indicated. (c) Snapshot of atom displacements of the $A_1$-like minor-$O_{Mg-S}^{Zn}$ mode, highly-localized near Zn (colored in blue).

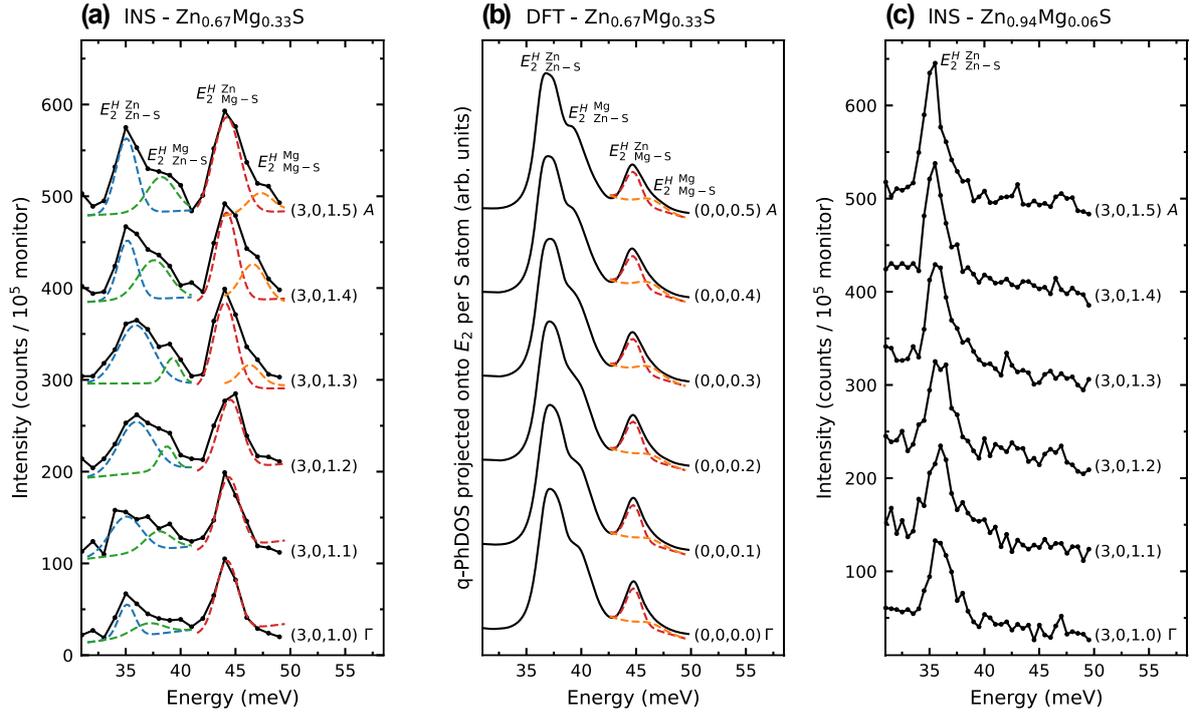

FIG. S7. $E_2^H$ phonon dispersion of wurtzite (w) $Zn_{1-x}Mg_xS$. (a) $E_2^H$ phonon dispersion of w-$Zn_{0.67}Mg_{0.33}S$ measured by INS along $\vec{c^*}$. The dual Zn-S and Mg-S signals (subscript) are deconvolved for clarity (colored dashed lines), distinguishing *homo* from *hetero* environments (superscript). (b) Corresponding $E_2^H$ phonon dispersion per S atom calculated a*b initio* (SIESTA DFT) using the w-$Zn_{0.67}Mg_{0.33}S$ special quasirandom structure shown in Fig. S4. (c) Corresponding $E_2^H$ phonon dispersion of w-$Zn_{0.94}Mg_{0.06}S$ measured by INS.

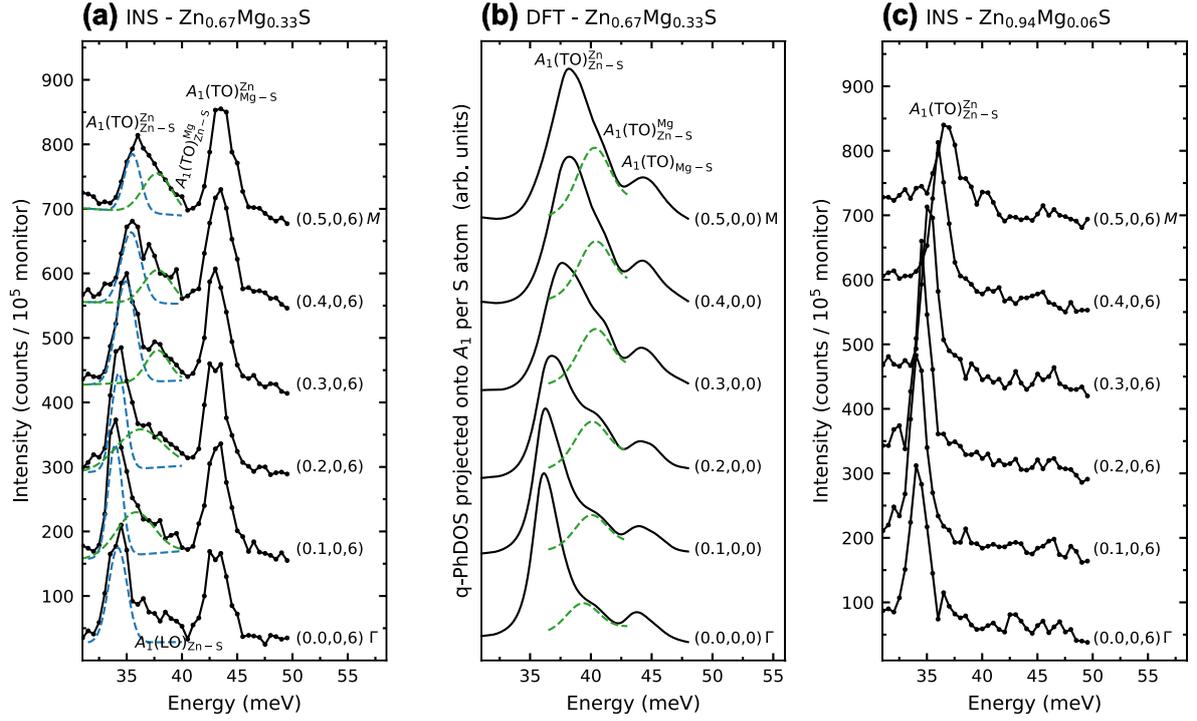

FIG. S8. $A_1(TO)$ phonon dispersion of wurtzite (w) $Zn_{1-x}Mg_xS$. (a) $A_1(TO)$ phonon dispersion of w-$Zn_{0.67}Mg_{0.33}$S measured by INS along $\vec{a^*}$. The dual Zn-S signal (subscript) is deconvolved for clarity (colored dashed lines), distinguishing *homo* from *hetero* environments (superscript). $A_1(LO)$ is activated specifically at the Brillouin zone centre, as indicated. (b) Corresponding $A_1(TO)$ phonon dispersion per S atom calculated a*b initio* (SIESTA DFT) using the w-$Zn_{0.67}Mg_{0.33}$S special quasirandom structure shown in Fig. S4. (c) Corresponding $A_1(TO)$ phonon dispersion of w-$Zn_{0.94}Mg_{0.06}$S measured by INS.

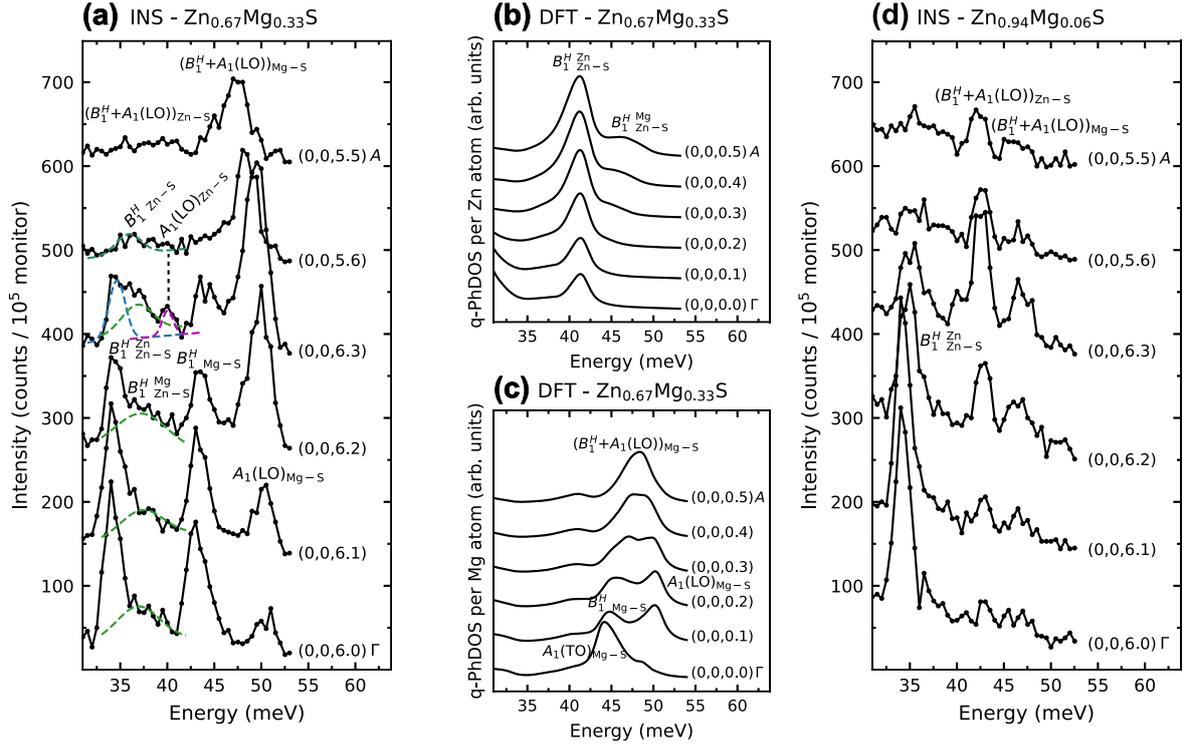

FIG. S9. Joined $(B_1^H + A1_{LO})$ phonon dispersions of wurtzite (w) $Zn_{1-x}Mg_xS$. (a) Joined $(B_1^H + A1_{LO})$ phonon dispersions of w-$Zn_{0.67}Mg_{0.33}S$ measured by INS along $\vec{c^*}$. The dual Zn-S $B_1^H$ signal is deconvolved for clarity (colored dashed lines), distinguishing *homo* from *hetero* environments (superscript). (b) and (c) Corresponding phonon dispersions per Zn and Mg atom, respectively, calculated a*b initio* (SIESTA DFT) using the w-$Zn_{0.67}Mg_{0.33}S$ special quasirandom structure shown in Fig. S4. (b) and (c) Corresponding phonon dispersions per Zn and Mg atom, respectively, calculated a*b initio* (SIESTA DFT) using the w-$Zn_{0.67}Mg_{0.33}S$ special quasirandom structure shown in Fig. S4. (d) Corresponding phonon dispersions of w-$Zn_{0.94}Mg_{0.06}S$ measured by INS.

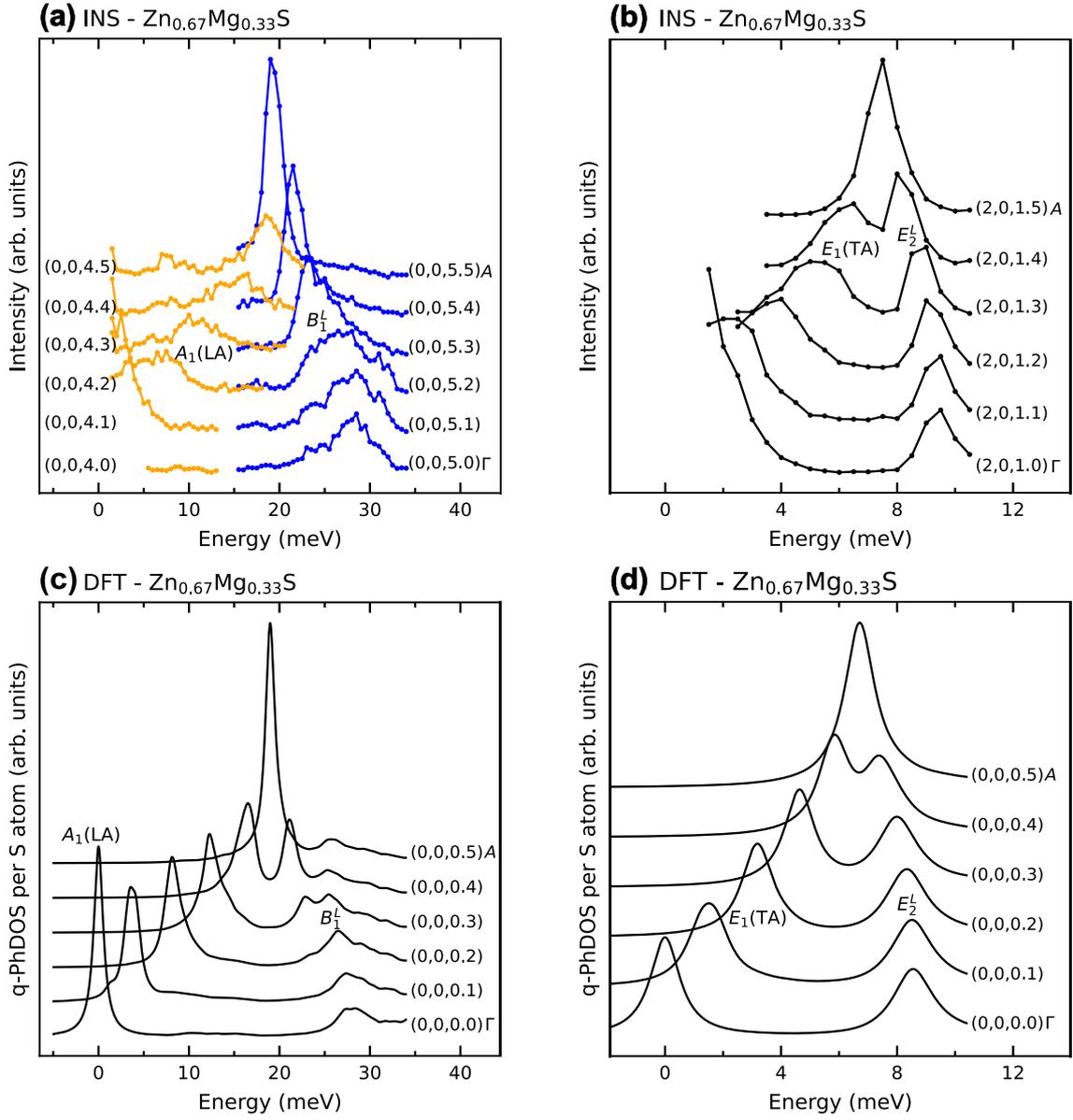

FIG. S10. Low-frequency phonon dispersions of wurtzite (w) $Zn_{0.67}Mg_{0.33}S$ along $\vec{c^*}$. (a) ($B_1^L$ (blue)+$LA$ (orange)) and (b) ($E_2^L+TA$) phonon dispersions of w-$Zn_{0.67}Mg_{0.33}S$ measured by INS along $\vec{c^*}$. (c) – (d) Corresponding *ab initio* (DFT) data obtained by using the relevant (right) special quasirandom structure shown in Fig. S4.

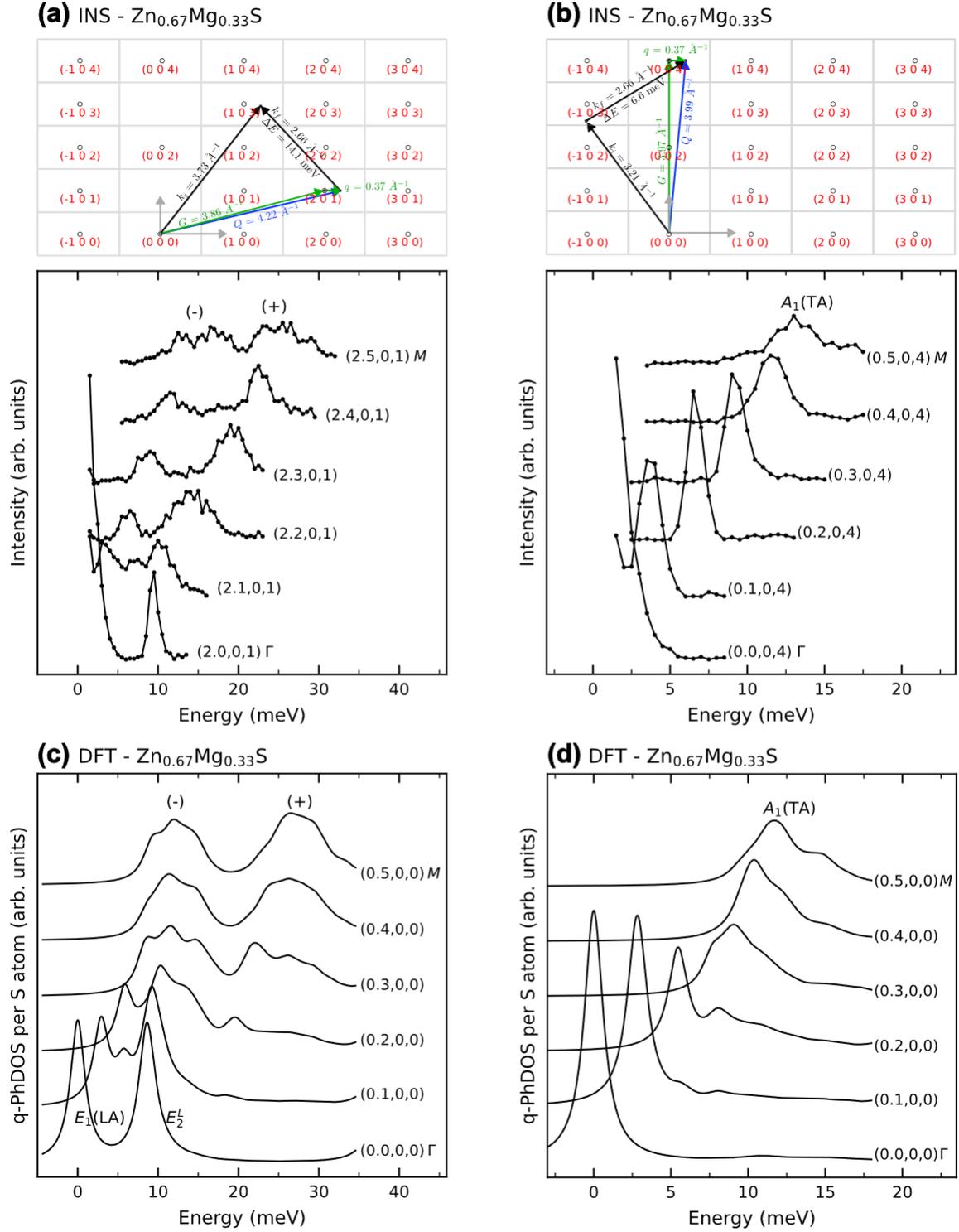

FIG. S11. Low-frequency phonon dispersions of wurtzite (w) Zn$_{0.67}$Mg$_{0.33}$S along $\vec{a^*}$. (a) A-phonon dispersions of w-Zn$_{0.67}$Mg$_{0.33}$S measured by INS along $\vec{a^*}$. (b) Same dispersion recorded specifically in the TA symmetry. The used scattering geometries are indicated in each case (insets) (c) – (d) Corresponding phonon dispersions per S atom Corresponding *ab initio* (DFT) data calculated *ab initio* (SIESTA DFT) using the w-Zn$_{0.67}$Mg$_{0.33}$S special quasirandom structure shown in Fig. S4.

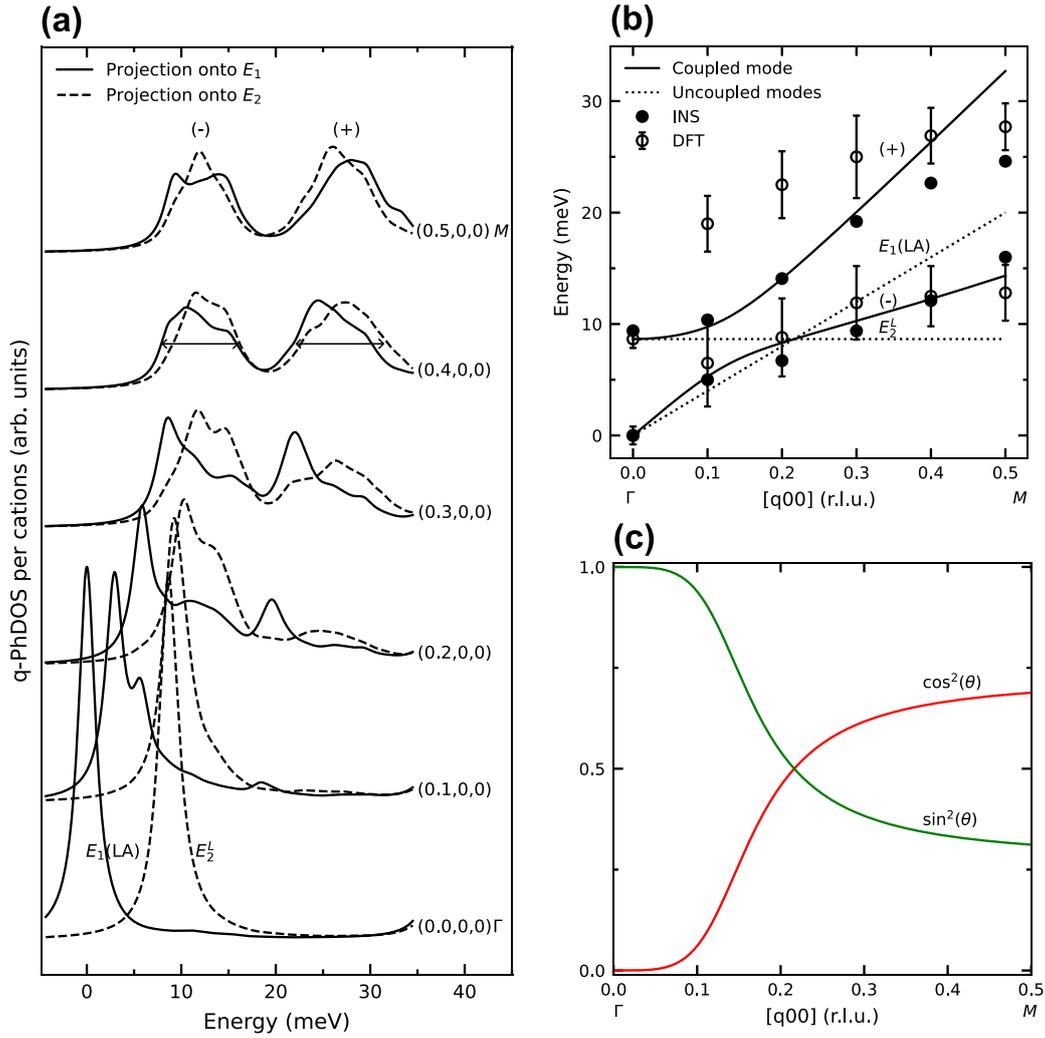

FIG. S12. $E_2^L - E_1(LA)$ coupling along $\vec{a^*}$. (a) *Ab initio* $\vec{q}$-projected $E_2^L$ (dotted curves) and $E_1(LA)$ (solid curves) PhDOS per cation (summed across Zn and Mg) behind the corresponding (±) coupled modes, completing the cumulated $E_2^L + E_1(LA)$ $\vec{q}$-projected PhDOS per anion (S) along $\vec{a^*}$ provided in Fig. S11c. (b) – (c) Corresponding model of two mechanically-coupled harmonic oscillators using $\omega'=40\times q$. (b) Dispersion of coupled modes (±) measured by INS (solid symbols, Fig. S11a) and predicted *ab initio* [hollow symbols, panel (a)]. The *ab initio* $\omega_\pm$ frequencies are marred by error bars estimated from the width at half mawimum of the *ab initio* lines in panel (a) (double arrows). (c) Relative contributions of $E_2^L$ ($cos^2\theta$) and $E_1(LA)$ ($sin^2\theta$) to the coupled mode (+) in their $\vec{q}$-dependence; and vice versa for the $E_2^L$ ($sin^2\theta$) and $E_1(LA)$ ($cos^2\theta$) contributions to the (coupled mode (−).